\documentclass[journal]{IEEEtran}
\IEEEoverridecommandlockouts
\usepackage{caption}

\usepackage[normalem]{ulem}
\usepackage{cite}
\usepackage{amsmath,amssymb,upgreek,fge,mathabx}
\usepackage{algorithmic}
\usepackage{textcomp}
\usepackage{xcolor}
\usepackage{cite}
\usepackage{url}
\usepackage{amssymb}
\usepackage{euscript}
\usepackage{amssymb}
\usepackage{pifont}

\usepackage{bbding}
\usepackage{graphicx}
\usepackage{multirow}
\usepackage{algorithm}
\usepackage{algorithmic}
\usepackage{ipa}
\usepackage[tone,extra,safe]{tipa}
\usepackage{changepage}
\usepackage{color}
\usepackage{subcaption}
\usepackage{makecell} 
\usepackage{todonotes}
\usepackage{scalerel}
\usepackage{tikz}
\usetikzlibrary{svg.path}
\usepackage{balance}
\usepackage{svg}
\usetikzlibrary{fit, positioning, shapes.geometric, fit, arrows.meta, calc, backgrounds}
\usepackage{pgfplots}
\pgfplotsset{compat=newest}
\usepackage{flushend}

\definecolor{darkgreen}{rgb}{0, 0.5, 0} 
\definecolor{lightpurple}{rgb}{0.7, 0.4, 1} 

\definecolor{orcidlogocol}{HTML}{A6CE39}
\tikzset{
orcidlogo/.pic={
\fill[orcidlogocol] svg{M256,128c0,70.7-57.3,128-128,128C57.3,256,0,198.7,0,128C0,57.3,57.3,0,128,0C198.7,0,256,57.3,256,128z};
\fill[white] svg{M86.3,186.2H70.9V79.1h15.4v48.4V186.2z}
svg{M108.9,79.1h41.6c39.6,0,57,28.3,57,53.6c0,27.5-21.5,53.6-56.8,53.6h-41.8V79.1z M124.3,172.4h24.5c34.9,0,42.9-26.5,42.9-39.7c0-21.5-13.7-39.7-43.7-39.7h-23.7V172.4z}
svg{M88.7,56.8c0,5.5-4.5,10.1-10.1,10.1c-5.6,0-10.1-4.6-10.1-10.1c0-5.6,4.5-10.1,10.1-10.1C84.2,46.7,88.7,51.3,88.7,56.8z};
}
}
\newcommand\orcidicon[1]{\href{https://orcid.org/#1}{\mbox{\scalerel*{
\begin{tikzpicture}[yscale=-1,transform shape]
\pic{orcidlogo};
\end{tikzpicture}
}{|}}}}

\newcommand{\zyy}[1]{\textcolor{blue}{#1}}

\usepackage[colorlinks=true,linkcolor=blue,citecolor=blue]{hyperref}

\def\BibTeX{{\rm B\kern-.05em{\sc i\kern-.025em b}\kern-.08em
T\kern-.1667em\lower.7ex\hbox{E}\kern-.125emX}}
\usepackage[most]{tcolorbox}

\begin{document} 
\author{
$ 
\text{Roberta Bucchignani}^{\orcidicon{0009-0006-2121-9580}, *}\ \IEEEmembership{Student Member, IEEE},$
$ 
\text{Furkan Bagci}^{\orcidicon{0009-0008-0473-5764}, *}\ \IEEEmembership{Student Member, IEEE},$
$
\text{Jalal Jalali}^{\orcidicon{0000-0002-3609-6775}}\ \IEEEmembership{Member, IEEE},$
\text{and}\ 
$ \text{Natasha Devroye}^{\orcidicon{0000-0002-1619-4095}}\ \IEEEmembership{Fellow, IEEE}.$ 
\thanks{
Roberta Bucchignani, Furkan Bagci, Jalal Jalali, and Natasha Devroye were affiliated with the Department of Electrical and Computer Engineering at the University of Illinois Chicago, Chicago, IL, USA. Roberta Bucchignani is currently with the Department of Electrical and Computer Engineering at Northeastern University, Boston, MA, USA. 
Jalal Jalali is also with the Wireless Communication Research Group, JuliaSpace LLC., Chicago, IL, USA.
The corresponding author is Furkan Bagci, fbagc@uic.edu.\\ 
$^*$ Bucchignani and Bagci are equal contribution student authors. 
}
\thanks{The work of N. Devroye was partially supported by NSF under award 2240532. The contents of this article are solely the responsibility of the authors and do not necessarily represent the official views of the NSF. 
The work of J. Jalali was supported in part by the Belgian American Educational Foundation (BAEF).
}
\thanks{\zyy{This work was submitted to the IEEE Transactions on Wireless Communications on July 26, 2026, and is currently under review.}}
}

\title{\huge Interpretability Analysis of End-to-End Learned CSI Feedback for Precoding-Oriented Massive MIMO}

\maketitle
\begin{abstract}
Recent end-to-end learning frameworks jointly optimize pilot design, channel state information (CSI) feedback, and downlink precoding, achieving strong performance under limited feedback overhead in massive multiple-input multiple-output (MIMO) networks. However, the internal representations and decision-making strategies learned by these architectures remain largely unexplored. This paper presents an interpretability analysis of a precoding-oriented end-to-end CSI feedback framework using explainable artificial intelligence (xAI) techniques. Building on a validated learned architecture, we investigate the structure of the learned pilots, latent feedback representations, and precoding vectors through low-dimensional visualization, density-based clustering, and cross-block representation analysis. The results reveal that the learned system undergoes a transition as the feedback constraint is relaxed. Under stringent feedback budgets, the encoder develops clustered latent representations and quantized precoding behavior. 
As additional feedback resources become available, the learned representations evolve into smooth geometry-preserving manifolds that retain physically meaningful channel characteristics such as dominant angular information, while enabling more adaptive precoding decisions. Furthermore, the learned pilot signals naturally form complementary sensing patterns aligned with the underlying channel statistics. The analysis also reveals a shift from gain-driven beamforming strategies under severe compression to increasingly interference-aware precoding as the feedback budget increases. These findings provide new insight into how end-to-end learned CSI feedback systems balance compression efficiency and communication performance and establish a general framework for interpreting learned communication architectures.
\end{abstract}
 

\section{Introduction}
\indent


\IEEEPARstart{M}{assive} MIMO has become a fundamental enabler for achieving significant increases in spectral efficiency, allowing base stations (BSs) equipped with large antenna arrays to simultaneously serve multiple users over shared time-frequency resources \cite{bjornson2016ten}. Traditionally, these systems operate efficiently under time-division duplexing (TDD), where channel reciprocity enables the BS to reuse uplink channel estimates for downlink precoding. However, many deployed cellular networks operate under frequency-division duplexing (FDD), where reciprocity cannot be exploited. In FDD systems, the user equipment (UE) must estimate the downlink channel and feed it back to the BS. As antenna arrays grow, the amount of channel state information (CSI) to be acquired and reported increases proportionally, resulting in a severe feedback overhead bottleneck that limits practical scalability \cite{bjornson2016ten}.
To mitigate this bottleneck, transmitting compressed CSI or task-relevant channel features over rate-limited feedback links becomes essential \cite{chen2024csipppnet}.

In current 5th-Generation New Radio (5gNR) systems, CSI feedback is configured through standardized mechanisms 
typically provided through predefined codebooks that describe candidate precoding matrices \cite{3gpp38331,3gpp38214,3gpp38211}. Simpler codebook options keep feedback overhead low but provide limited spatial resolution, while more advanced configurations improve spatial precision using increased feedback bits. This fundamental trade-off leads to overhead growth with antenna array size, creating severe scalability challenges in Massive MIMO deployments \cite{3gpp38214}.
To overcome the limitations of standardized codebooks, 
classical solutions for limited-feedback Massive MIMO systems have historically focused on two main categories: 
compressed sensing techniques \cite{Rao2014}, and codebook-based precoding selection methods, which allow users to report indices corresponding to preferred beams\cite{love2008overview,sohrabi2021deep}.

While advanced codebook-based strategies attempt to mathematically optimize precoder dictionaries, they inherently suffer from the same scalability bottlenecks as 5G codebooks when applied to massive arrays \cite{nair2020hybrid,castellanos2018channel,alkhateeb2015limited}.
Comparative studies have shown that under tight feedback constraints, directly feeding back quantized CSI often allows linear precoders to better exploit channel structure than selecting beams from limited codebooks, especially in propagation environments with sparse scattering where only a few dominant paths need to be characterized \cite{dietl2007linear,sohrabi2021deep}. 
Alternatively, compressed sensing approaches exploit this channel sparsity by acquiring only a limited number of measurements and reconstructing the channel through sparsity-promoting optimization, often operating in transformed domains where channel representations become sparse \cite{11404406}. Nevertheless, both of these classical approaches often rely on simplified channel assumptions and struggle to generalize across complex real-world propagation conditions. 


Motivated by these challenges, both academic and industrial research communities have explored alternative CSI compression strategies beyond fixed codebooks, specifically turning to learning-based feedback mechanisms that can adaptively capture task-relevant channel features \cite{11478444}.

\subsection{Prior Work: End-to-End Learned Feedback}
Deep learning has emerged as a powerful alternative for CSI compression and feedback, enabling end-to-end optimization across pilot design, user-side compression, and BS processing. Reconstruction-oriented autoencoder frameworks model CSI compression as learned source coding, where neural encoders map channel observations to compact latent representations that are transmitted and decoded to reconstruct the channel with minimal distortion, as first demonstrated by CsiNet \cite{wen2018deep} and improved in \cite{guo2022overview}. Building upon this idea, rate-distortion-driven approaches incorporate learned entropy models to explicitly control feedback overhead while minimizing reconstruction error, drawing inspiration from neural image compression frameworks such as the entropy bottleneck introduced in \cite{balle2018variational}. Methods such as DeepCMC \cite{mashhadi2021deepcmc} further integrate quantization and entropy coding within convolutional architectures, enabling explicit trade-offs between CSI reconstruction quality and feedback bit rates. However, these approaches primarily optimize reconstruction metrics rather than the ultimate downlink transmission objective.

To address this limitation, task-oriented learning approaches directly optimize precoding or system throughput rather than channel reconstruction accuracy. In multi-user settings, Sohrabi \emph{et al.} \cite{sohrabi2021deep} proposed an end-to-end framework where pilot design, user feedback encoding, and BS precoding are jointly learned to maximize downlink sum-rate without explicitly reconstructing channels. Their design directly accounts for inter-user interference and optimizes spectral efficiency, while further introducing training extensions to generalize the architecture across varying feedback budgets and different user counts. In single-user scenarios, implicit CSI feedback methods have been proposed where neural networks learn both what channel features should be transmitted and how they should be interpreted for beamforming, thereby reducing overhead compared with standardized codebooks while directly optimizing communication performance \cite{chen2022implicit}. In summary, deep learning-based feedback approaches consistently outperform classical compressed sensing methods at comparable overhead levels, with rate-distortion models providing explicit overhead control and task-oriented methods aligning training objectives with final communication performance.

Among recent advances, the framework proposed by Carpi \emph{et al.} \cite{carpi2023precoding} represents a significant step forward in precoding-oriented CSI feedback for multi-user FDD Massive MIMO systems under sparse limited-scattering channel models \cite{11456309}. Instead of treating channel estimation as a stand-alone module, their approach jointly designs downlink pilots, CSI compression, and BS precoding within an end-to-end learning architecture. User devices compress CSI over rate-limited links, and the BS directly generates precoders from compressed feedback without reconstructing channel matrices. Their architecture incorporates quantization and learned entropy models so that feedback overhead appears as a differentiable component of the optimization objective. Training simultaneously maximizes downlink sum-rate while penalizing feedback overhead, controlled by a trade-off parameter. Their results reveal a saturation effect in the rate-overhead trade-off, where increasing feedback beyond a certain point yields diminishing performance gains approaching perfect CSI performance. 

\subsection{Prior Work: Interpreting Learned Communication Blocks}
Despite the success of end-to-end learned feedback architectures, the internal mechanisms and latent representations of standard precoding-oriented CSI networks remain largely unexplored. While XAI has recently been introduced to wireless communications, early efforts have primarily treated the physical layer as a black box. For instance, in the context of radio resource management, Guo \cite{guo2020explainable} proposes an interpretability framework that seeks to map neural network decision-making to physical Key Performance Indicators (KPIs), such as spectral efficiency and interference, across both physical (PHY) and medium access control (MAC) layers. To achieve this, their methodology employs surrogate models and feature-importance techniques (e.g., LIME) which approximate the complex neural architecture with simpler, localized models. 
However, as Sun \emph{et al.} \cite{sun2025advancing} note, while general-purpose interpretability methods (e.g., SHAP and LIME) are effective for higher-layer network management or basic feature attribution, they fall short when applied to the physical layer. Complementing this perspective, Xiao \emph{et al.} \cite{xiao2026explainable} highlight the critical need for customized XAI tailored to communication-specific structures, as standard tools struggle to capture the complex, non-linear spatial transformations inherent in physical layer signal processing. By relying on generic ``post-hoc" feature attribution, these methods treat the network as an input-output black box, leaving the internal encoding of physical information unexplored.
To bypass this opacity, emerging research trends have shifted away from standard black-box architectures toward models that are interpretable by design. For instance, Gao \emph{et al.} \cite{gao2026bayesian} integrate variational Bayesian inference into FDD Massive MIMO to create a probabilistic, interpretable framework. Similarly, recent advancements propose replacing standard precoding-oriented CSI networks with Vector Quantized Variational Autoencoders (VQ-VAE) to explicitly force the learning of discrete codebooks that correlate with the underlying CSI \cite{chen2026precoding}. Concurrently, other approaches are shifting towards multi-modal foundational models, with physics-guided self-supervised pretraining to learn transferable, physics-aware representations \cite{yazdnian2026multimodal}.
While these architectural shifts are effective, the internal encoding within standard, deterministic continuous autoencoders remains an open question. Although these autoencoders remain the backbone of most end-to-end precoding research, the decoding of their internal spatial representations and their operational transitions across varying overhead regimes remains an unaddressed challenge. To address this problem, we propose a specialized framework that moves beyond standard feature-importance XAI, directly analyzing the geometric and operational behavior of the network’s latent space to uncover implicit beamforming strategies.

\subsection{Contributions and Paper Outline}

Motivated by the growing adoption of end-to-end learning for CSI feedback and precoding, this work investigates how such systems organize, compress, and utilize channel information under different feedback constraints. While prior studies have primarily focused on communication performance, the internal representations and decision-making processes learned by these architectures remain largely unexplored. To address this gap, we develop an XAI-based framework specifically designed for interpreting end-to-end learned CSI feedback systems. 
{While we adopt the state-of-the-art precoding-oriented architecture from \cite{carpi2023precoding} as a representative case study, our XAI methodology is framework-agnostic and broadly applicable to other end-to-end physical layer architectures.}

The main contributions of this work are summarized as follows:
\begin{itemize}
\item We develop a comprehensive XAI-based analysis framework for end-to-end learned CSI feedback systems, enabling the systematic study of learned pilot signals, latent CSI representations, and precoding vectors.
\item We investigate the geometric and structural properties of the learned representations using visualization, clustering, and cross-block analysis techniques. The results reveal a transition from coarse discrete compression behavior under strict feedback constraints to geometry-aware continuous representations with additional feedback. 
\item We demonstrate that the learned latent and precoding spaces preserve physically meaningful channel characteristics, particularly angular information, despite the absence of any explicit supervision on such parameters.
\item We provide new insights into the operational strategies learned by end-to-end precoding-oriented systems, showing how the architecture dynamically shifts between gain-driven and interference-aware transmission behaviors. 
\end{itemize}

The remainder of this paper is organized as follows. Section~\ref{section_2} presents the massive MIMO system model and problem formulation. Section~\ref{section_3} details the end-to-end learned CSI feedback framework. Section~\ref{section_4} introduces the XAI-based analysis methodology to study the learned representations and communication strategies. Section~\ref{section_5} provides an application and interpretability analysis. Finally, Section~\ref{section_6} concludes the paper and discusses future research directions.

\subsection{Notation}
\noindent Ordinary scalar quantities are represented by italic lowercase symbols (e.g., $x$), vector quantities by bold lowercase symbols (e.g., $\mathbf{x}$), and matrix quantities by bold uppercase symbols (e.g., $\mathbf{X}$). The Euclidean norm of a vector $\mathbf{x}$ is denoted by $\|\mathbf{x}\|$, while $\mathrm{Tr}(\mathbf{X})$ represents the trace of matrix $\mathbf{X}$. The conjugate transpose (Hermitian transpose) of $\mathbf{X}$ is expressed as $\mathbf{X}^{H}$. The notation $\mathcal{CN}(\mu,\sigma^{2})$ refers to a circularly symmetric complex Gaussian random variable with mean $\mu$ and variance $\sigma^{2}$. Furthermore, $\mathcal{U}[a,b]$ denotes the uniform distribution defined over the interval $[a,b]$.
The notation used throughout this paper is summarized in Table~\ref{table:notation_updated}.

\begin{table}[!t]
\caption{Summary of Main Notation}
\label{table:notation_updated}
\centering
\setlength{\tabcolsep}{3pt}
\renewcommand{\arraystretch}{1.15}
\begin{tabular}{|l|p{7.25cm}|}
\hline
\textbf{Symbol} & \textbf{Description} \\
\hline
$N_t$ & Number of BS transmit antennas \\
$K$ & Number of users \\
$L$ & Pilot sequence length \\
$L_p$ & Number of propagation paths \\
$P$ & Total BS transmit power budget \\
$\mathbf{h}_k$ & Downlink channel vector of user $k$ \\
$\hat{\mathbf{h}}_k$ & Reconstructed channel estimate of user $k$ \\
$\tilde{\mathbf{X}}$ & Learned downlink pilot matrix \\
$\tilde{\mathbf{x}}_{\ell}$ & $\ell$-th pilot vector (column of $\tilde{\mathbf{X}}$) \\
$\tilde{\mathbf{y}}_k$ & Pilot observations received by user $k$ \\
$\mathbf{z}_k$ & Additive Gaussian noise vector during pilot reception \\
$\mathbf{x}$ & Transmitted downlink signal vector \\
$\mathbf{s}$ & Vector of transmitted data symbols \\
$s_k$ & Data symbol intended for user $k$ \\
$\mathbf{V}$ & Downlink precoding matrix \\
$\mathbf{v}_k$ & Precoding vector associated with user $k$ \\
$\mathrm{SINR}_k$ & Signal-to-interference-plus-noise ratio of user $k$ \\
$R_k$ & Achievable rate of user $k$ \\
$R$ & Network sum-rate \\
$\mathcal{F}$ & User-side CSI feedback function \\
$\mathcal{G}$ & BS-side processing function \\
$f_{\theta}$ & Neural encoder at the user side \\
$g_{\phi}$ & Neural network at the BS side \\
$q(\cdot)$ & Uniform scalar quantizer \\
$c_{\psi}$ & Entropy encoder \\
$c_{\psi}^{-1}$ & Entropy decoder \\
$\mathbf{t}_k$ & Latent representation generated by user $k$ \\
$\tilde{\mathbf{t}}_k$ & Pseudo-quantized latent representation used during training \\
$\bar{\mathbf{t}}_k$ & Quantized latent representation \\
$\mathbf{b}_k$ & Compressed CSI feedback message of user $k$ \\
$N_b$ & Latent representation dimension \\
$\mathbf{u}_k$ & Uniform noise used to approximate quantization during training \\
$\mathcal{O}$ & Feedback overhead metric \\
$\mathcal{R}$ & Sum-rate utility metric \\
$\mathcal{D}$ & Channel reconstruction distortion metric \\
$\lambda$ & Overhead--performance tradeoff coefficient \\
$\gamma$ & Distortion weighting coefficient \\
$\alpha_{\ell,k}$ & Complex gain of path $\ell$ for user $k$ \\
$\beta_{\ell,k}$ & Angle of departure (AoD) of path $\ell$ for user $k$ \\
$\mathbf{a}_t(\beta)$ & Transmit array steering vector \\
\hline
\end{tabular}
\end{table}

\section{System Model and Problem Formulation}\label{section_2}
\begin{figure}[!t]
\centering
\includegraphics[width=8.5cm,trim=4 4 4 4,clip]{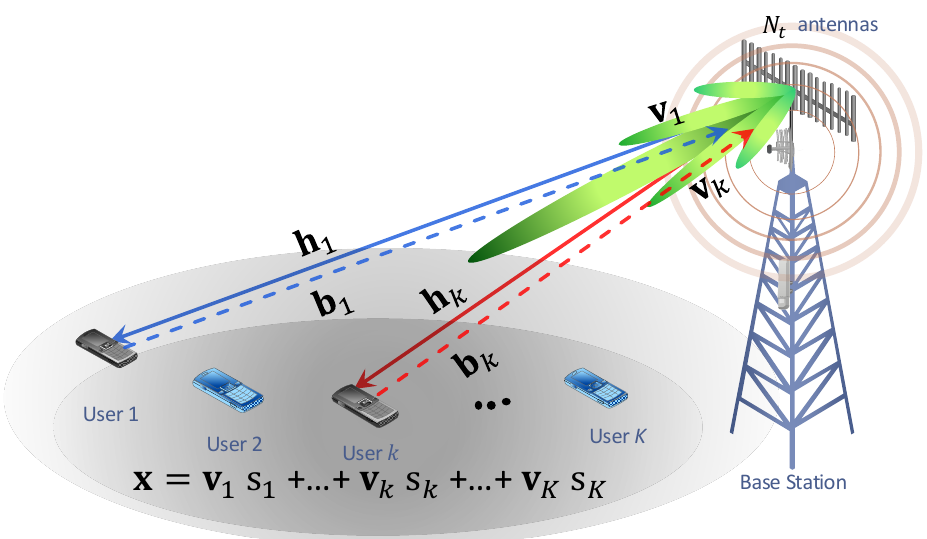}
\caption{Considered multi-user massive MIMO downlink system. A BS equipped with $N_t$ transmit antennas simultaneously serves $K$ single-antenna users. The transmitted signal is formed as $\mathbf{x}=\sum_{k=1}^{K}\mathbf{v}_k s_k$, where $\mathbf{v}_k$ denotes the precoding vector for the symbol $s_k$ intended for user $k$. After propagation through the downlink channels $\mathbf{h}_1,\ldots,\mathbf{h}_K$, each user receives its signal, estimates its channel, and transmits a compressed binary CSI feedback message $\mathbf{b}_1,\ldots,\mathbf{b}_K$ to the BS. The BS reconstructs the channel information from the received feedback and computes the downlink precoding vectors for subsequent data transmission.}
\label{fig:1}
\end{figure}

We consider the same channel model and framework as in \cite{sohrabi2021deep, carpi2023precoding}: a cellular massive MIMO deployment composed of a multi-antenna BS serving geographically distributed users. 
As shown in Fig.~\ref{fig:1}, the network contains one BS equipped with $N_t$ transmit antennas. Within the cell, $K$ single-antenna users are simultaneously served across the network.
The signal observed by user $k$ can be expressed as:
\begin{equation}
y_{k} = {\mathbf{h}}_{k}^{H} {\mathbf {x}}
+ z_{k}, 
\label{eq1}
\end{equation}
where ${\mathbf{h}}_{k}\in\mathbb{C}^{N_t}$ denotes the channel vector between the serving BS and the user. 
The transmitted downlink signal is denoted by ${\mathbf{x}}\in\mathbb{C}^{N_t}$. 
The term $z_{k}\sim\mathcal{CN}(0,\sigma^2)$ models additive white Gaussian noise.
The BS applies linear precoding for downlink transmission, 
and the transmitted signal can be written as:
\begin{equation}
{\mathbf {x}} = \sum _{k=1}^{K} {\mathbf {v}}_{k} s_{k} = {\mathbf {V}} {\mathbf {s}},
\label{eq2}
\end{equation}
where ${\mathbf{v}}_{k}\in\mathbb{C}^{N_t}$ denotes the precoding vector associated with the $k$-th user, and $s_{k}\in\mathbb{C}$ represents the transmitted symbol intended for that user. 
The matrix ${\mathbf V}\in\mathbb{C}^{N_t\times K}$ collects the precoders for all users. 
The transmit power is constrained such that $\text{Tr}({\mathbf V}{\mathbf V}^H)\leq P$, with $P$ as the total available transmit power, while the data symbols satisfy $\mathbb{E}[{\mathbf s}{\mathbf s}^H]=\mathbf I$.
The signal received by user $k$ can be decomposed into the desired signal component, intra-cell interference, and noise:
\begin{align}
y_{k} & = 
{\mathbf {h}}_{k}^{H} 
{\mathbf {v}}_{k}s_{k} + 
\sum_{l \neq k} {\mathbf {h}}_{k}^{H} 
{\mathbf {v}}_{l} s_{l} + z_{k}.
\label{eq3}
\end{align}
From this expression, the signal-to-interference-plus-noise ratio (SINR) experienced by user $k$ is defined as:
\begin{equation}
\text {SINR}_{k} = 
\frac { | {\mathbf {h}}_{k}^{H} 
{\mathbf {v}}_{k}|^{2} }{ \sum _{l\neq k} | {\mathbf {h}}_{k}^{H} {\mathbf {v}}_{l}|^{2} + \sigma^2}, 
\label{eq4}
\end{equation}
where $\sum _{l\neq k} | {\mathbf {h}}_{k}^{H} {\mathbf {v}}_{l}|^{2}$ represents the intra-cell interference power.
Consequently, the achievable rate for user $k$ becomes:
\begin{equation}
R_{k} = 
\log _{2} 
\left ({{ 1 + \text {SINR}_{k} }}\right). 
\label{eq6}
\end{equation}
The network is composed of $K$ users, and hence the overall network throughput is characterized by: 
\begin{equation}
R = \sum _{k=1}^{K} R_{k}.
\label{eq7}
\end{equation}
The precoding matrices ${\mathbf V}$ are optimized jointly to maximize the network sum-rate in \eqref{eq7}. 
To design these beamforming vectors, the BS requires accurate knowledge of the downlink channel conditions. 
Specifically, the BS transmits known reference signals,  referred to as pilots, allowing users to estimate their respective channels. Users then compress these observations and transmit the CSI feedback over the uplink, allowing the BS to compute the final precoding matrices.


\begin{figure*}[t]
    \centering
    \includegraphics[trim={3.5cm 3.0cm 3.5cm 2.5cm},clip,width=1\textwidth]{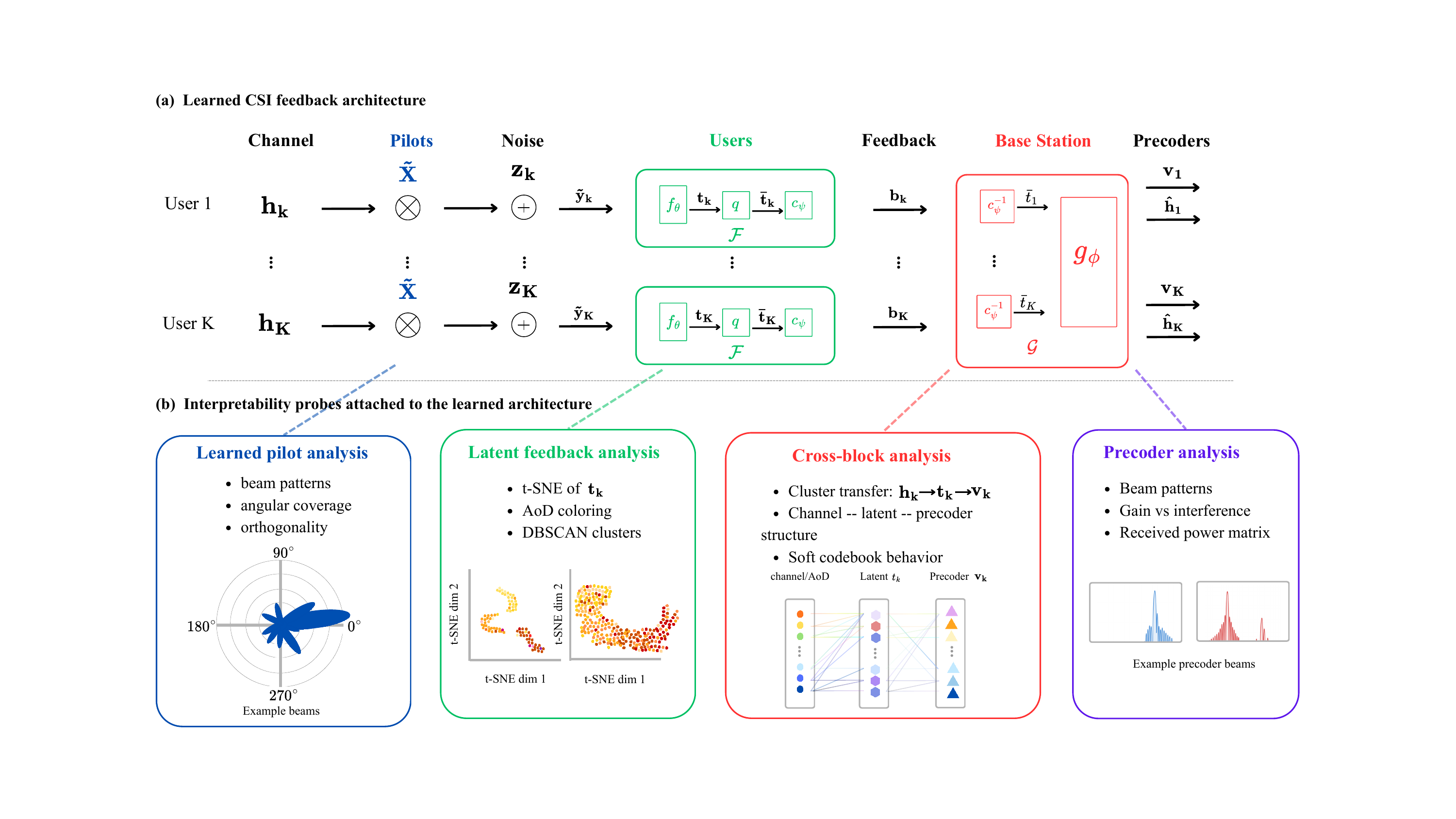}
    \caption{Overview of the considered end-to-end communication architecture. The pilot signals, highlighted in blue, are optimized during training. The user-side feedback module, shown in green and denoted by $\mathcal{F}$, consists of the neural encoder $f_{\theta}$, the quantization block $q$, and the entropy coding unit $c_{\psi}$. At the BS, the processing function $\mathcal{G}$ includes the entropy decoder $c_{\psi}^{-1}$ and the neural network $g_{\phi}$. Depending on the selected design objective, the BS outputs either the downlink precoding vectors $\mathbf{v}_k$ in the precoding-oriented framework or the reconstructed channel estimates $\hat{\mathbf{h}}_k$ in conventional CSI recovery approaches. (b) Interpretability probes attached to key internal variables to reveal what the system senses, preserves, and learns at each stage.}
    \label{fig:architecture}
\end{figure*}

\subsection{Pilot-Based Channel Estimation}

Let $\tilde{\mathbf{X}}\in\mathbb{C}^{N_t\times L}$ denote the pilot matrix transmitted by the BS during the downlink training phase, where $L$ represents the pilot sequence length. The observation received by the $k$-th user is given by:
\begin{align}
{\mathbf {\tilde {y}}}_{k} 
& = {\mathbf {h}}_{k}^{H} {\mathbf {\tilde {X}}} + {\mathbf {z}}_{k},
\label{eq8}
\end{align}
where ${\mathbf z}_{k} \in \mathbb{C}^{1 \times L}$ represents the additive noise vector, distributed as $\mathcal{CN}(\mathbf{0},\sigma^2 \mathbf{I}_L)$.
Each pilot symbol must satisfy the instantaneous power constraint $\|\tilde{\mathbf{x}}_{\ell}\|_2^2 \leq P$, 
where $\tilde{\mathbf{x}}_{\ell}$ denotes the $\ell$-th column of pilot matrix $\tilde{\mathbf{X}}$.

\subsection{Compressed CSI Feedback Mechanism}
The CSI dimensionality increases rapidly with the number of BS antennas and users, making direct feedback of raw channel observations impractical. 
Therefore, users process the received pilot signals locally to extract and compress the most relevant channel features. The compressed feedback message generated by the $k$-th user is defined as:
\begin{align}
{\mathbf {b}}_{k} 
& = \mathcal {F}_{k}( {\mathbf {\tilde {y}}}_{k}), 
\label{eq9}
\end{align}
where $\mathcal{F}_{k}$ denotes the local compression function and ${\mathbf b}_{k}$ represents the compressed feedback transmitted to the BS\footnote{For analytical clarity, we assume that the uplink feedback links are noiseless, meaning that the compressed feedback messages ${\mathbf b}_{k}$ are received by the BS without distortion. 
This standard assumption in CSI feedback literature \cite{8744528,9931713} enables a direct evaluation of the compression mechanism's efficiency.} over a rate-limited link.

\subsection{Precoding for downlink transmission}
Upon receiving the feedback messages $(\mathbf{b}_{1},\ldots,\mathbf{b}_{K})$ from all users, the BS generates the downlink precoding matrix and, optionally, the reconstructed channel matrix as:
\begin{align} 
(\mathbf{V},\hat{\mathbf{H}})=
\mathcal{G}
\left(
\mathbf{b}_{1},\ldots,\mathbf{b}_{K}
\right),
\label{eq10}
\end{align}
where $\mathcal{G}$ denotes the BS processing function. While $\mathcal{G}$ can output both representations, this paper focuses on a precoding-oriented architecture where the primary output is the matrix $\mathbf{V}$, as the sum-rate objective in \eqref{eq7} depends directly on the downlink precoders. The auxiliary reconstructed channel $\hat{\mathbf{H}}$ is generated solely to provide a baseline comparison with conventional CSI reconstruction-based approaches~\cite{dahlman20205g}. 



\section{The end-to-end learned system}\label{section_3}
We now present the learned framework for precoding-oriented CSI feedback. As illustrated in Fig.~\ref{fig:architecture}, the proposed architecture replaces conventional pilot design, CSI feedback, and BS processing methods \cite{9296555,sohrabi2021deep} with trainable neural network modules. The framework also incorporates learned quantization and entropy coding mechanisms, enabling compact CSI representations under limited feedback resources.

The user-side processing, shown in green, consists of a neural encoder $f_{\theta}$, a quantization block $q$, and an entropy coder $c_{\psi}$, which together implement the feedback function $\mathcal{F}$. The BS-side processing, shown in red, is composed of the entropy decoder $c_{\psi}^{-1}$ and the neural network $g_{\phi}$, which together realize the processing function $\mathcal{G}$. Unlike conventional CSI feedback schemes that first reconstruct the channel and subsequently compute the precoders, the proposed framework directly learns a feedback representation tailored to the precoding task. Consequently, the BS processing block can generate the downlink precoding matrix directly from the received feedback messages. Depending on the considered architecture, the output of $\mathcal{G}$ consists of either the precoding vectors $\{\mathbf{v}_{1},\ldots,\mathbf{v}_{K}\}$ or the reconstructed channel estimates $\{\hat{\mathbf{h}}_{1},\ldots,\hat{\mathbf{h}}_{K}\}$. 
By jointly training all components, the entire communication chain is optimized end-to-end rather than designing pilot transmission, CSI feedback, and precoding as separate processing stages. 

\subsection{Learned Downlink Pilots}

We begin with the pilot transmission stage used for downlink channel estimation. The received pilot observations at the $k$-th user are defined in \eqref{eq8}. Each pilot matrix $\tilde{\mathbf{X}}$ is modeled as the output of a single-layer fully connected neural network with linear activation and zero bias, which is learned jointly with the remaining blocks of the architecture. The input to this pilot-generation block {during training} is given by all channel realizations ${\mathbf{h}}_{k}$. {At inference time, the pilots remain the fixed learned pilots.} 
During training, the transmit power constraint is enforced by requiring each pilot symbol to satisfy $\|\tilde{\mathbf{x}}_{\ell}\|_{2}^{2}=P$ for $\ell=\{1,\dots,L\}$. Furthermore, additive Gaussian noise $\mathbf{z}_k$ is incorporated into the received pilot observations to model receiver noise. Once generated, the pilots are transmitted through the channel and the corresponding noisy observations are collected by the users. 

\subsection{Learned Uplink Feedback Scheme}

The user-side feedback module is responsible for generating the compressed CSI information that is transmitted to the BS through the uplink feedback channel. As illustrated in Fig.~\ref{fig:architecture}, all users employ the same feedback function $\mathcal{F}$, while operating on their own received pilot observations. This shared architecture is motivated by the fact that users experience channel realizations drawn from the same underlying statistical model, allowing a common feedback strategy to generalize across different terminals.

The feedback function $\mathcal{F}$ consists of three stages: a neural encoder $f_{\theta}$, a quantization block $q$, and an entropy coder $c_{\psi}$. The neural encoder extracts a compact representation from the received pilot observations. For the $k$-th user, the latent representation is given by $\mathbf{t}_{k}
=
f_{\theta}(\tilde{\mathbf{y}}_{k})$,
where $\mathbf{t}_{k}\in\mathbb{R}^{N_b}$ and $N_b$ denotes the dimension of the latent space. In this work, $f_{\theta}$ is implemented as a fully connected DNN whose parameters are learned jointly with the remaining blocks.

The latent representation is subsequently processed by the quantization block, which applies uniform scalar quantization to each element of $\mathbf{t}_{k}$. The resulting quantized vector is:
$\bar{\mathbf{t}}_{k}
=
q(\mathbf{t}_{k})$.
Since quantization is not differentiable, it cannot be directly incorporated into gradient-based optimization. To address this issue, the quantization operation is replaced during training by additive uniform noise. Specifically, a noise vector $\mathbf{u}_{k}$ is generated with independent components following:
\begin{align}
u_{k,i} \sim \mathcal{U}[-0.5,0.5],
\qquad i=\{1,\ldots,N_b\},
\end{align}
and a pseudo-quantized representation is formed as $\tilde{\mathbf{t}}_{k}
=
\mathbf{t}_{k}
+
\mathbf{u}_{k}$.
This differentiable approximation enables gradients to propagate through the quantization stage during training.\\
{Finally, the entropy coder $c_{\psi}$ is employed}
to efficiently convert $\bar{\mathbf{t}}_{k}$ into a binary feedback message:
\begin{align}
\mathbf{b}_{k}
=
c_{\psi}
(\bar{\mathbf{t}}_{k}).
\end{align}
The resulting bit stream $\mathbf{b}_{k}$ constitutes the CSI feedback message transmitted by the $k$-th user to the BS. 

\subsection{Learned Precoding Computation}

After the compressed CSI messages are received, the BS processes them to generate the downlink precoding matrix. As shown in Fig.~\ref{fig:architecture} and described in \eqref{eq10}, the BS processing block is denoted by $\mathcal{G}$ and consists of the entropy decoder $c_{\psi}^{-1}$ followed by the neural network $g_{\phi}$.

During inference, the entropy decoder losslessly reconstructs the quantized latent representation associated with each user according to
\begin{align}
\bar{\mathbf{t}}_{k}
=
c_{\psi}^{-1}(\mathbf{b}_{k}),
\qquad 
k=\{1,\ldots,K\}.
\end{align}
During training, however, the entropy coding and decoding operations are bypassed to preserve differentiability. Instead, the pseudo-quantized latent vectors $\tilde{\mathbf{t}}_{k}$ are directly provided as inputs to the BS neural network.

In its most general form, the BS processing function may produce both the downlink precoding matrix and reconstructed channel estimates, i.e., \eqref{eq10}, which is
$
(\mathbf{V},\hat{\mathbf{H}})
=
\mathcal{G}
\left(
\mathbf{b}_{1},\ldots,\mathbf{b}_{K}
\right).
$
However, the focus of this work is on the precoding-oriented architecture, where only the precoding matrix is generated as
\begin{align}
\mathbf{V}
=
g_{\phi}
\left(
\tilde{\mathbf{t}}_{1},
\ldots,
\tilde{\mathbf{t}}_{K}
\right),
\end{align}
during training, while the decoded latent vectors $\bar{\mathbf{t}}_{k}$ are used during inference. 
The resulting matrix ${\mathbf V}$ must satisfy the transmit power constraint, enforced by imposing $\mathrm{Tr}({\mathbf V}{\mathbf V}^{H})=P$.
This design enables the BS to map the received CSI feedback directly to the downlink precoder without explicitly reconstructing the full channel matrices.

\subsection{Differentiable Feedback Overhead Optimization}

The proposed framework explicitly incorporates the feedback overhead into the training objective. A key challenge, however, is that both quantization and entropy coding are non-differentiable operations, which prevents the direct application of gradient-based learning. 
{As discussed, the non-differentiable quantization and entropy coding stages are relaxed during training via additive uniform noise and a learned entropy model, a technique widely adopted in learned compression systems \cite{18,19,theis2017lossy,balle2018variational}.}

To estimate the feedback overhead in a differentiable manner, the latent variables are modeled through a learned probabilistic distribution parameterized by $\psi$. Following \cite{balle2018variational}, a fully factorized density model is adopted, where each latent component is represented by a zero-mean Gaussian distribution whose variance is learned. The resulting continuous density associated with $\tilde{\mathbf{t}}_{k}$ acts as a relaxation of the discrete probability mass function of the quantized vector $\bar{\mathbf{t}}_{k}$ and  the differential entropy of $\tilde{\mathbf{t}}_{k}$ provides an accurate approximation of the entropy of $\bar{\mathbf{t}}_{k}$ and therefore of the average coding rate produced by the entropy coder. This entropy estimate is incorporated into the loss function and serves as a differentiable measure of the CSI feedback overhead.

During deployment, the artificial noise is removed and the latent representation generated by the encoder is processed through the actual quantizer and entropy coder before transmission. In this way, the system is trained using a differentiable surrogate while operating with true quantization and entropy coding at test time.


\subsection{Metrics for the Feedback–Performance Tradeoff}

The purpose of the proposed architecture is to optimize the tradeoff between CSI feedback overhead and communication performance. To train the system with deep learning, we therefore define a differentiable loss function that combines these objectives. The loss to be minimized is written as~\cite{renzo2019smart}:
\begin{equation}
\mathcal {L}(\theta , \psi , \phi )
=
\mathcal {O}
-
\lambda \mathcal {R}
+
\gamma \mathcal {D},
\label{loss_function}
\end{equation}
where $\mathcal{O}$ denotes the feedback overhead, $\mathcal{R}$ represents the achievable sum-rate, and $\mathcal{D}$ quantifies the channel reconstruction distortion. The non-negative coefficients $\lambda$ and $\gamma$ determine the relative importance of communication performance and reconstruction accuracy with respect to the feedback cost. 

The formulation in \eqref{loss_function} encompasses multiple CSI feedback design paradigms. For instance, conventional rate--distortion optimization is obtained by setting $\lambda=0$, whereas precoding-oriented CSI feedback corresponds to $\gamma=0$. Since the objective of this work is the direct generation of downlink precoders rather than channel reconstruction, we focus exclusively on the latter case. Consequently, the operational loss function reduces to:
\begin{equation}
\mathcal {L}(\theta , \psi , \phi ) = \mathcal {O} - \lambda \mathcal {R}.
\label{eq:operational_loss}
\end{equation}
During training, all parameters $(\theta,\psi,\phi)$ are optimized jointly in an end-to-end manner. The individual components of the loss function are described next.

\paragraph*{i) Performance}
The system performance is measured through the achievable sum-rate across all users, recalling \eqref{eq6} and \eqref{eq7}:
\begin{equation}
\begin{split}
\mathcal{R}(\theta, \psi, \phi) &= \sum_{k=1}^{K} \mathcal{R}_{k}(\theta, \psi, \phi) \\
&= \sum_{k=1}^{K} \mathbb{E}_{\mathbf{h}_{k}, \mathbf{u}_{k}, \mathbf{z}_{k}} \left[ \log_{2} \left( 1 + \text{SINR}_{k} \right) \right],
\end{split}
\end{equation}
with $\text{SINR}_{k}$ given in \eqref{eq4}. 
The precoding vectors used in this expression are generated by the learned BS processing block.

\paragraph*{ii) Overhead}
The second term captures the feedback cost required to deliver the compressed CSI messages ${\mathbf b}_{k}$ from the users to the BS. This system overhead aggregates the empirical entropy of the pseudo-quantized latent vectors across all users as:
\begin{equation}
\begin{split}
\mathcal{O}(\theta, \psi) &= \sum_{k=1}^{K} \mathcal{O}_{k}(\theta, \psi) \\
&= \sum_{k=1}^{K} \mathbb{E}_{\mathbf{h}_{k}, \mathbf{u}_{k}, \mathbf{z}_{k}} \big[ -\log_{2} p_{\mathbf{\tilde{t}}}(\mathbf{\tilde{t}}_{k}; \psi) \big],
\end{split}
\end{equation}
where $p_{\mathbf{\tilde{t}}}(\cdot;\psi)$ denotes the learned probability density of the pseudo-quantized latent representation.

\section{XAI-Based Analysis Framework}
\label{section_4}

{Unlike traditional evaluations that focus solely on final performance metrics (e.g., achievable sum-rate), our goal is to characterize the internal representations and uncover the underlying strategies learned by the neural networks. To achieve this, we propose a systematic analysis framework employing a complementary suite of XAI tools
to examine the geometric and operational flow of the communication pipeline, as detailed in the following subsections.}

\subsection{Low-Dimensional Visualization via t-SNE}
\label{sec_tsne}

To gain geometric insight into the learned high-dimensional representations, we employ t-distributed Stochastic Neighbor Embedding (t-SNE) \cite{maaten2008visualizing}. This technique maps a collection of high-dimensional feature vectors into a low-dimensional embedding space while preserving local neighborhood relationships between nearby samples. 
More specifically, t-SNE constructs joint probability distributions to measure pairwise similarities in both the original space (using Gaussian distributions) and the embedded space (using a Student's t-distribution). The mapping is optimized by minimizing the Kullback--Leibler (KL) divergence between these two distributions~\cite{9151281}.
While t-SNE is effective at revealing complex nonlinear structures that may not be captured by linear dimensionality reduction techniques, e.g., PCA, its visualizations must be interpreted with caution. Since t-SNE primarily preserves local neighborhood relationships, distances between well-separated clusters in the low-dimensional embedding do not necessarily correspond to their true separations in the original high-dimensional space~\cite{9729516}. In this work, t-SNE is employed as a qualitative visualization tool to examine the geometric organization of the learned latent representations and precoding vectors under different feedback constraints. Specifically, it enables us to investigate whether the learned mappings form smooth, geometry-preserving manifolds or instead concentrate around a finite set of representative solutions driven by aggressive feedback compression.

\subsection{Semantic Interpretation via Attribute Coloring}
\label{sec_coloring}
To interpret the physical meaning of the learned representations, we associate each sample with communication-relevant attributes, such as channel gain, SINR, user location, and angular direction. By coloring the t-SNE embeddings based on these attributes, we identify correlations between the abstract latent space and actual physical system parameters. This reveals whether the neural network intrinsically organizes its representations around meaningful spatial and propagation features.

\subsection{Clustering via Density-Based Methods}\label{sec_DBSCAN}
To quantitatively validate the structures observed in the visualizations, we apply density-based clustering using DBSCAN \cite{ester1996density}. Unlike centroid-based methods (e.g., k-means), DBSCAN does not require prior knowledge of the number of clusters and is particularly suited for identifying irregularly shaped structures in high-dimensional embeddings. Given a set of representations, DBSCAN identifies clusters as dense regions separated by low-density areas. Specifically, a point is classified as a \emph{core point} if at least \texttt{MinPts} neighboring samples lie within a distance threshold $\varepsilon$. Clusters are formed by recursively connecting density-reachable core points, while isolated samples are treated as outliers. 

We apply DBSCAN to the learned latent representations to determine whether the encoder partitions the channel manifold into discrete semantic regions. Such clusters can be interpreted as learned decision regions, 
providing insight into whether the feedback mechanism behaves as a continuous mapping or more like a  discrete, ``codebook-like'' strategy under different feedback budgets.

\subsection{Cross-Block Structural Interpretation}
\label{sec_blocks}

To understand how information flows through the architecture, we examine the structural relationships across different pipeline stages: the original channel space, the latent CSI representations, and the decoded precoding vectors. Rather than analyzing each component in isolation, we transfer clustering labels and visual patterns across these spaces to identify consistent operational behaviors. This cross-block analysis clarifies how the encoder compresses spatial channel information and how the BS decoder translates this latent data into explicit beamforming decisions, which will be extensively demonstrated in the following section.


\section{Interpretability Analysis of the Learned CSI Feedback Architecture}
\label{section_5}

In this section, we provide an interpretation of the end-to-end learned precoding-oriented CSI feedback framework proposed in \cite{carpi2023precoding} 
using the XAI-based tools detailed in Section~\ref{section_4}. 
We first present the overall system setup and analyze the fundamental overhead--performance tradeoff achieved by the learned framework. We then investigate the geometric structure and operational behavior of the learned pilots, latent CSI feedback representations, and downlink precoders. Finally, we provide system-level interpretations revealing how the architecture dynamically transitions between different operational regimes as the feedback constraint varies. 

\subsection{Overall System Setup and Overhead--Performance Tradeoff}
\subsubsection{Channel, Architecture, and Training Setup}
The downlink channel for each user is generated using a geometric multipath model with $L_p=2$ propagation paths. Specifically, the channel vector for the $k$-th user is constructed as:
\begin{equation}
\mathbf{h}_k
=
\frac{1}{\sqrt{L_p}}
\sum_{\ell=1}^{L_p}
\alpha_{\ell,k}\,\mathbf{a}_t(\beta_{\ell,k}),
\end{equation}
where $\alpha_{\ell,k}\sim\mathcal{CN}(0,1)$ is the complex path gain and $\mathbf{a}_t(\beta)$ is the uniform linear array steering vector given by:
\begin{equation}
\mathbf{a}_t(\beta)
=
\left[
1,\,
e^{j\frac{2\pi f_c d}{c}\sin(\beta)},
\,\ldots,\,
e^{j\frac{2\pi f_c d}{c}(N_t-1)\sin(\beta)}
\right]^T.
\end{equation}
Here, $\beta$ denotes the angle of departure (AoD), $d$ is the antenna spacing, $f_c$ is the carrier frequency, and $c$ is the speed of light. Unless otherwise stated, the simulation and training parameters are summarized in Table~\ref{tab:sim_params}. The setup follows the geometric channel model and training configuration adopted in~\cite{carpi2023precoding}. 



\begin{table}[!t]
\centering
\caption{Simulation and Training Parameters}
\label{tab:sim_params}
\begin{tabular}{ll}
\hline
\textbf{Parameter} & \textbf{Value} \\
\hline
Number of BS antennas $N_t$ & 64 \\
Number of users $K$ & 2 \\
Number of channel paths $L_p$ & 2 \\
Carrier frequency $f_c$ & 3.5 GHz \\
Antenna spacing $d$ &  $c/(2 f_c)$ \\
Pilot length $L$ & 8 \\
Latent dimension $N_b$ & 16 \\
Training batch size & 1024 \\
Number of training iterations & $10^{6}$ \\
Learning rate & $10^{-3}$ \\
Optimizer & Adam \\
Transmit power constraint $P$ & 1 \\
Training SNR & 10 dB \\
AoD distribution & $\beta_{\ell,k}\sim\mathcal{U}[-30^\circ, 30^\circ]$ \\
Path gain distribution & $\alpha_{\ell,k}\sim\mathcal{CN}(0,1)$ \\
Channel model & Geometric multipath model \\
Quantization method & Uniform scalar quantization \\
Entropy model & Differentiable entropy bottleneck \\
Training relaxation noise &
$\mathbf{u}_k \sim \mathcal{U}[-0.5,0.5]$ \\
\hline
\end{tabular}
\end{table}

\begin{table}[!t]
\centering
\caption{Fully-Connected Deep Neural Network Architecture.}
\setlength{\tabcolsep}{4pt}
\renewcommand{\arraystretch}{1.5}
\begin{tabular}{llll}
\hline
\textbf{Component} & \textbf{In} & \textbf{Hidden} & \textbf{Out} \\
\hline
$f_\theta$ & $2L{=}16$ & $1024{\to}2048{\to}256$ & $N_b{=}16$ \\
$g_\phi$   & $KN_b{=}32$ & $1024{\to}512{\to}512{\to}256$ & $KN_t{\cdot}2{=}256$ \\
\hline
\multicolumn{4}{l}{\small BatchNorm + ReLU per hidden layer; linear output.}\\
\hline
\end{tabular}
\label{tab:architecture}
\end{table}

The neural network architectures are summarized in  Table~\ref{tab:architecture}. The user encoder $f_\theta$ takes the real and imaginary parts of the received pilots as input 
and maps them to a latent vector $\mathbf{t}_k \in \mathbb{R}^{N_b}$. 
The BS decoder $g_\phi$ takes the concatenated feedback from all $K$ users and directly outputs the precoding matrix.
All hidden layers use batch normalization followed by ReLU activation, with linear activation at the output. 
The tradeoff between overhead and performance is explored by varying $\lambda$ over the range $[0, 500]$. 

\subsubsection{Overhead--Performance Tradeoff Analysis}

To validate our implementation of the framework proposed in \cite{carpi2023precoding}, we benchmark the learned system against two ideal precoding schemes with perfect CSI at the transmitter: zero-forcing (ZF) and maximal-ratio transmission (MRT), which serve as upper and lower performance bounds, respectively.

\begin{figure}[t]
\centering
\resizebox{\columnwidth}{!}{
    \includegraphics[width=\columnwidth]{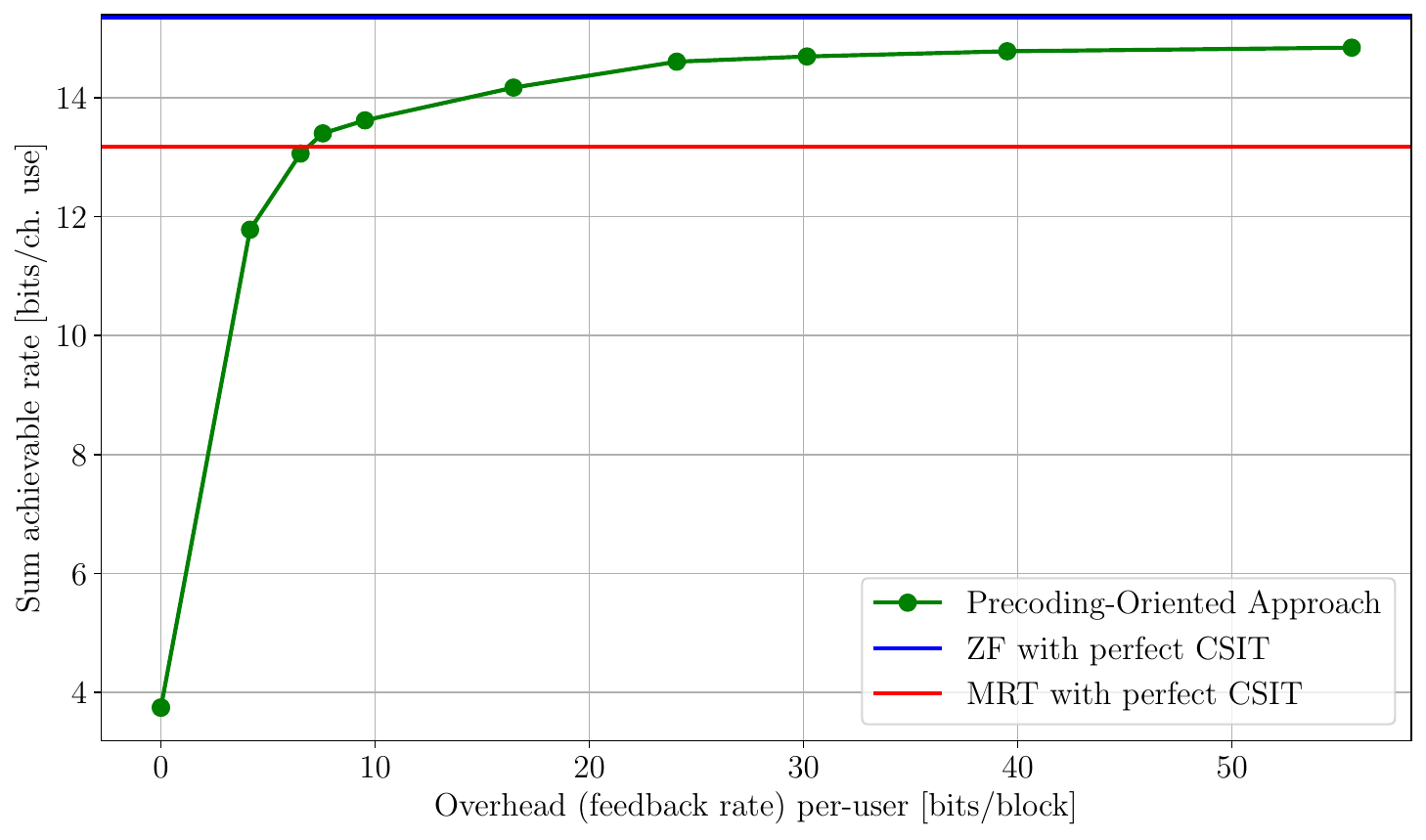}
}
\caption{Tradeoff between feedback overhead and sum-rate. ZF and MRT with perfect CSI are shown as baselines.}
\label{fig:rate_overhead}
\end{figure}

Fig.~\ref{fig:rate_overhead} illustrates the tradeoff between feedback overhead $\mathcal{O}$ and achievable sum-rate $\mathcal{R}$ for different operating points obtained by varying $\lambda \in \{0.1, 0.5, 1, 2, 5, 10, 25, 50, 100, 200, 500\}$. 
Throughout the subsequent visualizations and comparative analyses, we use $\lambda=2$ and $\lambda=100$ as representative low- and high-overhead operating points, respectively.
The learned framework consistently outperforms MRT across all overhead regimes and gradually approaches the ZF upper bound as the feedback budget increases. The most significant gains are observed in the low-overhead region, indicating that the learned compression mechanism successfully extracts task-relevant channel information without reconstructing the full CSI. As the feedback overhead increases, the achievable rate begins to saturate.

Our interpretability conclusions are summarized as follows:

{
\renewcommand{\labelitemi}{$\triangleright$}
\begin{itemize}
    \setlength{\itemsep}{0pt}
    \setlength{\parskip}{0pt}
    \setlength{\parsep}{0pt}

    \item The learned framework outperforms MRT and approaches ZF as the feedback overhead increases.

    \item $\lambda=2$ and $\lambda=100$ represent the low- and high-overhead regimes used in the following analyses.
\end{itemize}
}

\begin{figure}[t]
\centering
\resizebox{\columnwidth}{!}{
    \includegraphics[trim={0.2cm 0.0cm 0.2cm 0.0cm},clip,width=\textwidth]{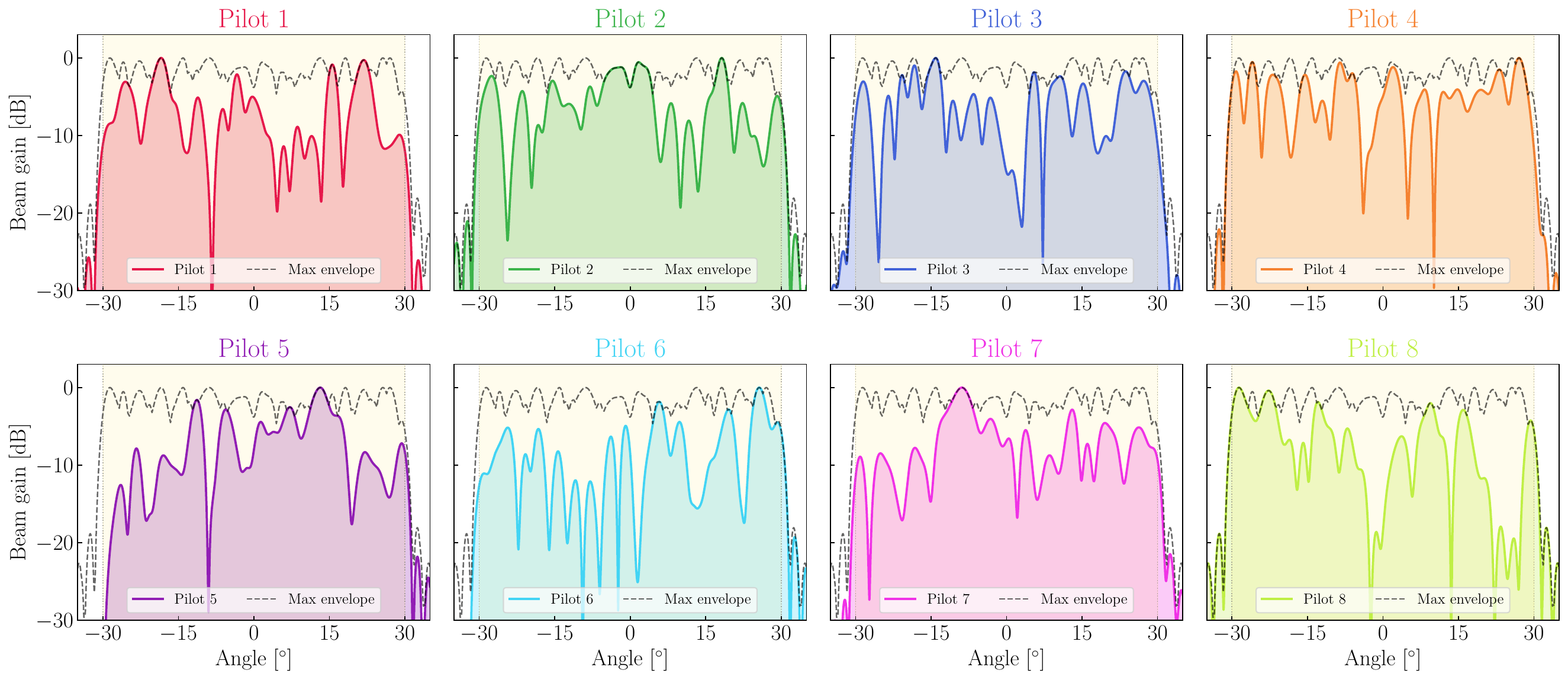}
}
\caption{Beampatterns of $L=8$ learned pilots for $\lambda=2$ ($N_t=64$). Colored curves show individual gains; the dashed black curve shows the maximum envelope, {where the max is taken over all 8 beampatterns.}}
\label{fig:pilots0}
\end{figure}

\begin{figure}[t]
\centering
\resizebox{\columnwidth}{!}{
    \includegraphics[trim={0.2cm 0.0cm 0.2cm 0.0cm},clip,width=\textwidth]{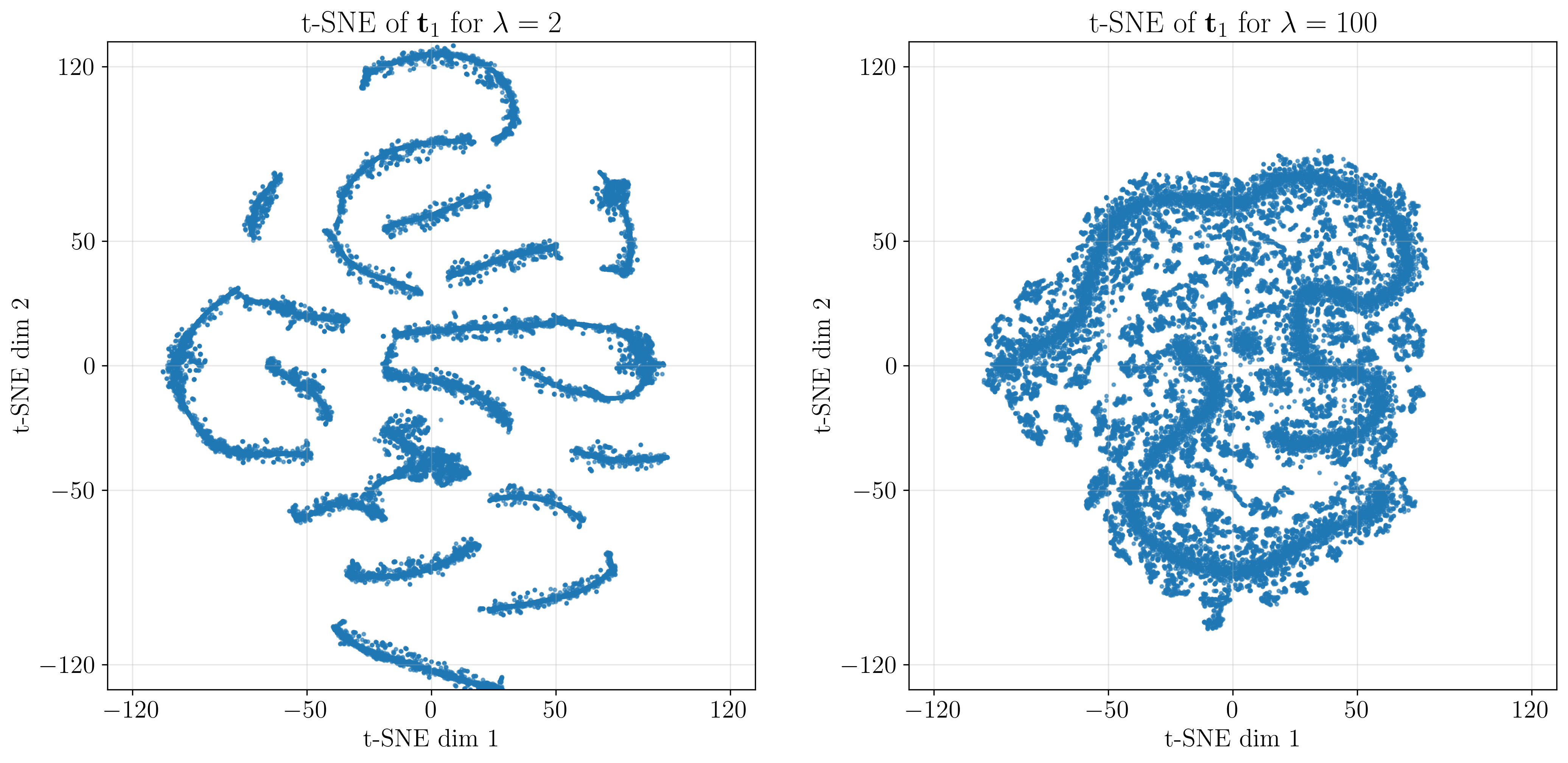}
}
\caption{t-SNE visualization of user-1 latent feedback vectors $\mathbf{t}_1$ for $\lambda=2$ (left) and $\lambda=100$ (right).}
\label{fig:tsne0}
\end{figure}

\begin{figure*}[t]
\centering
\resizebox{\textwidth}{!}{
    \includegraphics[trim={0.2cm 0.0cm 0.2cm 0.0cm},clip,width=\textwidth]{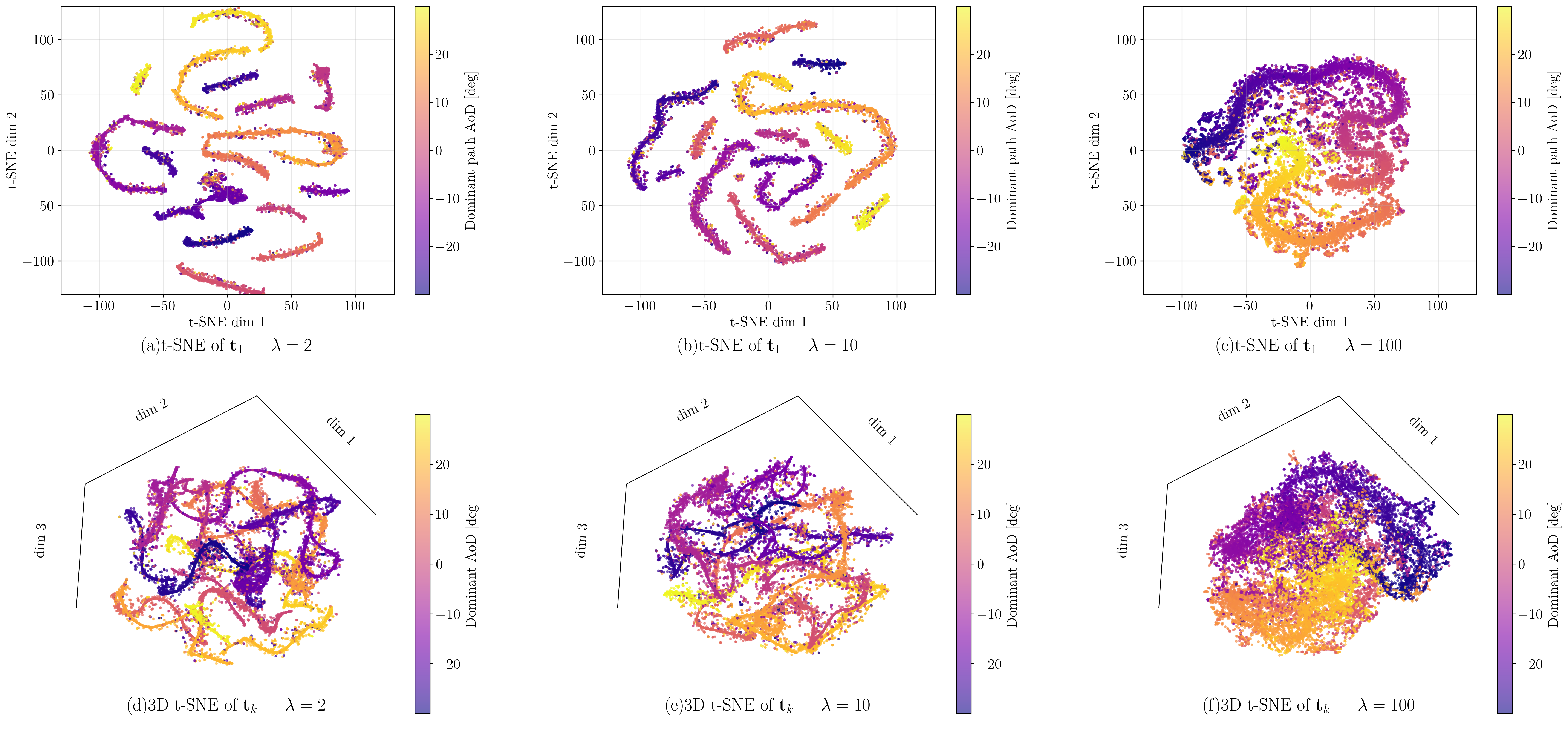}
}
\caption{2D and 3D t-SNE visualization of user-1 feedback vectors $\mathbf{t}_1$ colored by dominant path AoD. Top row: 2D t-SNE for (a) $\lambda=2$, (b) $\lambda=10$, and (c) $\lambda=100$. Bottom row: corresponding 3D t-SNE for (d) $\lambda=2$, (e) $\lambda=10$, and (f) $\lambda=100$.}
\label{fig:tsne_AoD}
\end{figure*}

\subsection{Interpretations of the Learned Pilots}

Moving beyond standard performance metrics, we investigate the learned pilot signals to understand how the architecture performs channel sensing under limited feedback constraints. Unlike conventional approaches that impose predefined beam structures or orthogonality constraints, the proposed framework jointly optimizes the pilots alongside the encoder and decoder using solely the end-task objective. 

\subsubsection{Learned Pilot Beam Patterns}

Fig.~\ref{fig:pilots0} illustrates the beampatterns of the learned pilot vectors. Although each pilot exhibits a multi-lobe pattern with noticeable nulls, these are naturally complemented by the others, creating a fairly uniform envelope over the AoD range $[-30^\circ, 30^\circ]$. 
This demonstrates that the pilots jointly and efficiently cover the relevant angular space. 

We present the figure for $\lambda=2$; however, the pilot patterns for $\lambda=100$ are visually similar, suggesting their design is largely insensitive to the rate-overhead tradeoff in this setup.  Furthermore, while not shown, the learned pilot matrix is near-orthogonal across both regimes, with a maximum inter-pilot correlation below 0.06. This confirms that the pilots provide largely independent spatial measurements, a property that emerges purely from end-to-end training without any explicit orthogonality constraint in the loss function. 
Finally, despite being randomly initialized over a wider angular domain, the learned pilots concentrate their coverage strictly within the $[-30^\circ, 30^\circ]$ AoD range, proving that the model implicitly adapts to the underlying channel statistics and focuses its sensing resources exactly where needed. 

Our interpretability conclusions are summarized as follows:

{
\renewcommand{\labelitemi}{$\triangleright$}
\begin{itemize}
\setlength{\itemsep}{0pt}
\setlength{\parskip}{0pt}
\setlength{\parsep}{0pt}

\item The learned pilots learned to jointly cover the channel AoD range $[-30^\circ,30^\circ]$ seen in training.

\item  The learned pilots are nearly orthogonal, providing complementary spatial measurements.

\end{itemize}
}


\subsection{Interpretations of the Learned CSI Feedback}

To understand how the encoder compresses and organizes channel information under different feedback constraints, we project the high-dimensional latent vectors $\mathbf{t}_{k}$ onto a two-dimensional plane using t-SNE.

\subsubsection{t-SNE Visualization of Latent Feedback Representations}

Following the low-dimensional visualization approach introduced in Section \ref{sec_tsne}, we apply t-SNE to project the latent feedback vectors $\mathbf{t}_1$ into two dimensions. Fig.~\ref{fig:tsne0} shows the resulting embeddings for two representative values of $\lambda$. In the low-overhead regime ($\lambda=2$), the latent space exhibits a fragmented structure composed of multiple well-separated clusters. 
This behavior indicates that the encoder maps channel realizations to a limited set of representative regions, resembling a quantized, codebook-like encoding strategy.
Such clustering reflects strong compression constraints, where preserving fine-grained channel geometry is secondary to reducing feedback overhead.

In contrast, in the high-overhead regime ($\lambda=100$), the latent space becomes significantly smoother and more continuous. The embeddings form coherent manifolds where neighboring points correspond to similar channel realizations. This structure suggests that the encoder inherently preserves geometric relationships within the channel space when sufficient feedback capacity is available. 


\subsubsection{Dominant AoD Preservation and Channel Geometry}


Following the semantic interpretation method in Section \ref{sec_coloring}, we color the t-SNE embeddings according to the dominant path AoD. To verify whether this geometric organization captures physically meaningful features, we color each embedded point according to its dominant AoD.

Fig.~\ref{fig:tsne_AoD} reveals a clear color ordering across all operating regimes: nearby points in the latent space exhibit similar dominant AoDs. The encoder preserves angular proximity in its learned representation, identifying the dominant propagation direction as one of the most relevant features for precoding; this emerges naturally from the end-to-end objective, as the model is never explicitly trained to estimate or preserve AoD information. However, the macro-structure of this space evolves alongside the feedback tradeoff parameter $\lambda$. In the low-overhead regime ($\lambda = 2$), although the AoD values remain relatively consistent within each cluster, transitions are abrupt. The discrete partitioning indicates that the encoder behaves as a coarse, codebook-based quantization mechanism. Fine angular distinctions are lost, prioritizing overhead reduction over geometric fidelity.

At intermediate values ($\lambda = 10$), the space begins to exhibit partial continuity. The previously isolated clusters elongate into locally smooth trajectories: the encoder starts to preserve relative angular differences between channels, representing a compromise between compression efficiency and geometric fidelity. 

In the high-overhead regime ($\lambda = 100$), the embeddings form coherent, continuous manifolds with smooth AoD gradients, indicating that neighboring latent vectors correspond to channels with similar propagation directions. This demonstrates that, when sufficient feedback resources are available, the encoder learns a geometry-aware representation that perfectly preserves the dominant spatial characteristics of the wireless channel. This continuous structural behavior perfectly aligns with the precoding requirements of highly adaptive spatial beam steering and interference mitigation. 

Overall, these interpretations demonstrate that the learned encoder dynamically adjusts its representation strategy according to the available feedback budget. 

\subsubsection{Clustering Behavior}

\begin{figure}[t]
\centering
\resizebox{\columnwidth}{!}{
    \includegraphics[trim={0.2cm 0.0cm 0.2cm 0.0cm},clip,width=\textwidth]{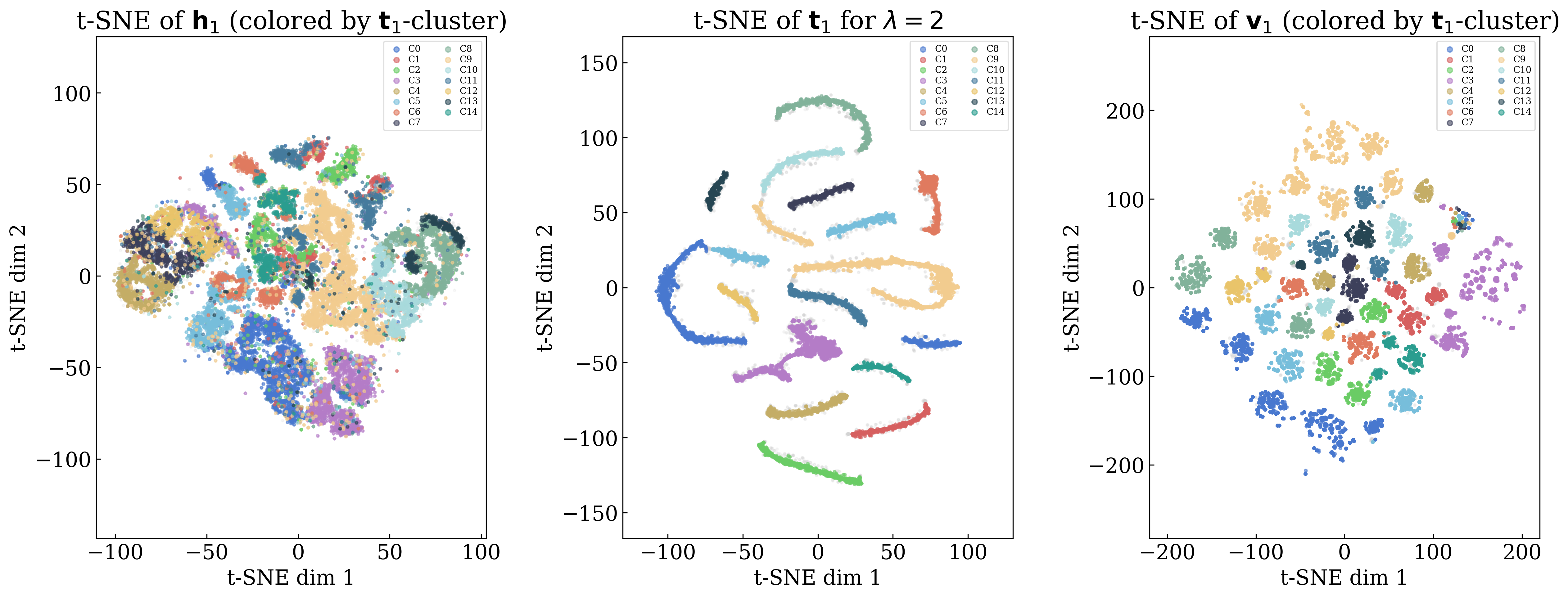}
}
\caption{t-SNE of $\mathbf{h}_1$, $\mathbf{t}_1$, and $\mathbf{v}_1$ (left to right) for $\lambda=2$. Latent clusters (via DBSCAN on $\mathbf{t}_1$) are transferred across spaces.}
\label{fig:mapping0}
\end{figure}

Following the density-based clustering and cross-block analysis methodologies detailed in Sections \ref{sec_DBSCAN} and \ref{sec_blocks}, we apply DBSCAN \cite{ester1996density} to the latent representations $\mathbf{t}_1$ to probe the internal structure learned by the encoder.  To ensure robustness and avoid manual tuning, the distance threshold $\varepsilon$ is selected via a systematic sweep, retaining only values that yield stable clustering across neighboring settings. We then transfer the resulting cluster labels to the channel space $\mathbf{h}_1$ and precoder space $\mathbf{v}_1$ to analyze how the learned partitions map to physical channels and generated beamforming decisions.

For $\lambda=2$, DBSCAN identifies 15 distinct clusters. As shown in Fig.~\ref{fig:mapping0} (middle), these clusters form well-separated, arc-shaped structures in the latent space. When the same cluster labels are visualized in the channel space $\mathbf{h}_1$ (left), they appear more intermixed, suggesting that the encoder transforms the channel representations into a more separable latent structure. Consistent with Fig.~\ref{fig:tsne_AoD}, the clusters also correspond to distinct dominant-AoD regions, indicating that the encoder performs a coarse angular partitioning of the channel space.

Visualizing these labels in the precoder space $\mathbf{v}_1$ (right) reveals that this structure is only partially preserved. While samples from the same latent cluster produce related precoding directions, they do not collapse into a single compact group. 
We refer to this as soft codebook-like behavior: each latent cluster corresponds to a family of related precoding solutions rather than a single fixed beamforming vector.
This variability reflects residual channel differences within each cluster, including secondary-path angles and path gains, allowing the decoder to make fine-grained adaptations while preserving the coarse angular structure learned by the encoder.

Our interpretability conclusions are summarized as follows:

{
\renewcommand{\labelitemi}{$\triangleright$}
\begin{itemize}
    \setlength{\itemsep}{0pt}
    \setlength{\parskip}{0pt}
    \setlength{\parsep}{0pt}

    \item Our proposed interpretability visualization tool, t-SNE, suggests that the latent representation transitions from separated clusters under limited feedback to a smooth, continuous manifold as the feedback budget increases.

    \item By coloring the visualized learned  variables with respect to relevant communication parameters, we see that the dominant-AoD information is preserved across operating regimes, showing that the encoder organizes the feedback around physically meaningful channel geometry.

    \item Under limited feedback, the encoder forms coarse angular partitions that compress channels with similar dominant directions into common latent regions.

    \item Cross-block analysis, based on following clusters across the different learned spaces, shows that these latent regions correspond to families of related precoding solutions, resulting in soft codebook-like behavior rather than a single fixed precoder per cluster. 
\end{itemize}
}

\subsection{Interpretations of the Learned Precoding Vectors} 

We next analyze how the learned feedback representations are transformed into precoding decisions. We first examine the geometry of the precoder space using t-SNE and dominant-AoD coloring, and then study the resulting beam patterns and interference-management behavior.

\subsubsection{t-SNE Visualization of Decoded Precoding Vectors}
\begin{figure}[t]
\centering
\resizebox{\columnwidth}{!}{
    \includegraphics[trim={0.2cm 0.0cm 0.2cm 0.0cm},clip,width=\textwidth]{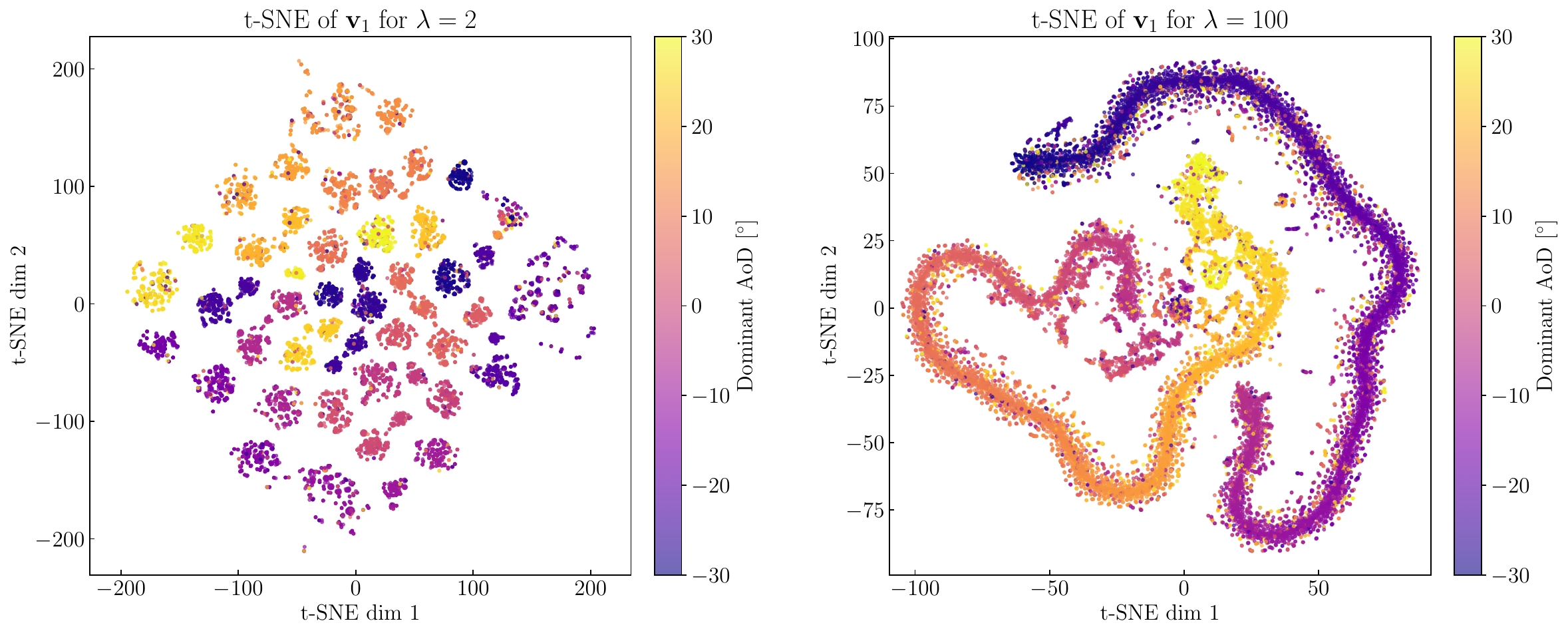}
}
\caption{t-SNE visualization of user-1 decoded precoding vectors $\mathbf{v}_1$ for $\lambda=2$ (left) and $\lambda=100$ (right).}
\label{fig:tsne_pre}
\end{figure}

Following the visualization and semantic-coloring methods introduced in Sections \ref{sec_tsne} and \ref{sec_coloring}, we apply t-SNE to the precoding vectors $\mathbf{v}_1$ and color the embeddings according to the dominant AoD. As shown in Fig.~\ref{fig:tsne_pre}, the precoder space undergoes a geometric transition that closely mirrors the latent representations.

\begin{figure*}[t]
\centering
\resizebox{\textwidth}{!}{
    \includegraphics[trim={0.2cm 0.0cm 0.2cm 0.0cm},clip,width=\textwidth]{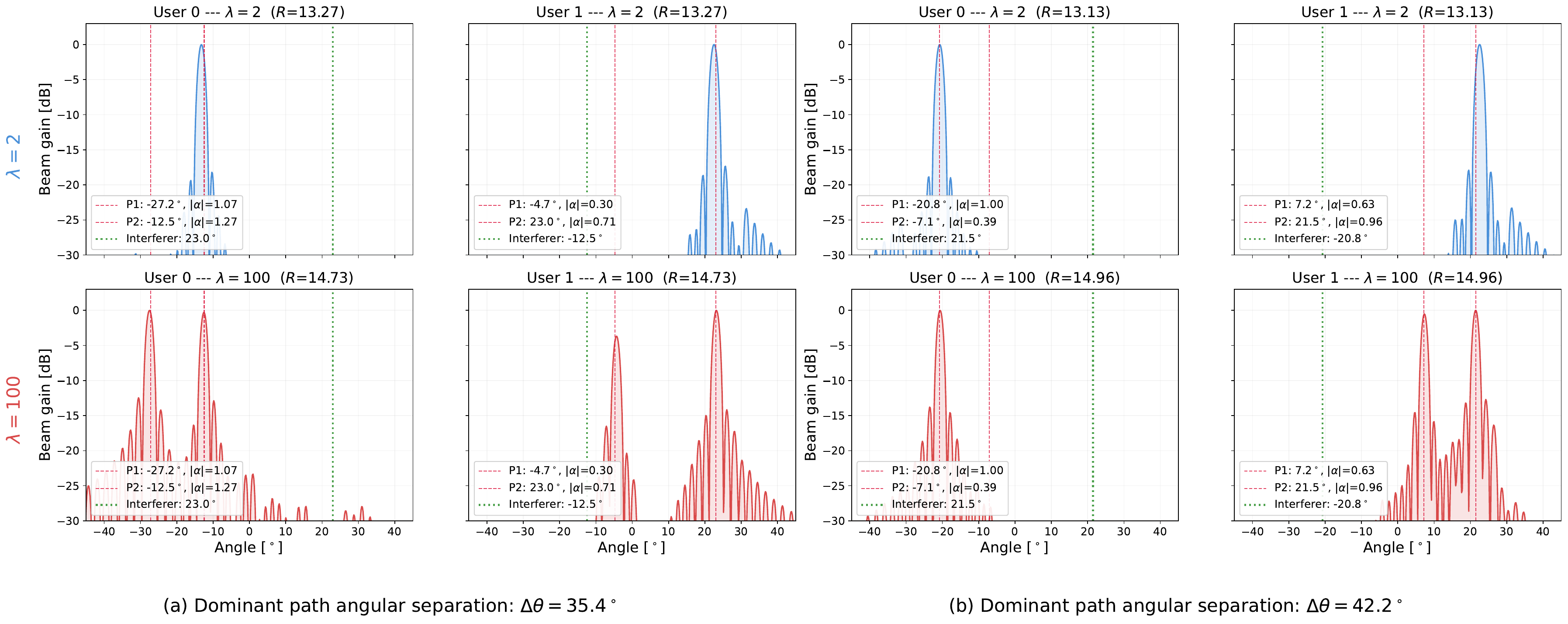}
}
  \caption{Beam patterns of the learned precoders for average channel
    realizations under $\lambda = 2$ (top row) and $\lambda = 100$ (bottom row).
    Red dashed lines indicate the user's own multipath AoDs with corresponding
    path gains $|\alpha|$; green dotted lines indicate the interfering user's
    dominant AoD. P1 and P2 denote path 1 and 2 of the stated user.}
\label{fig:beams0}
\end{figure*}

The resulting embeddings reveal a clear structural difference across operating regimes. In the low-overhead regime ($\lambda = 2$), the precoder space forms a small number of well-separated clusters: this indicates that the decoder effectively learns a discrete set of beamforming patterns, exhibiting a codebook-like transmission strategy. Multiple channel realizations are mapped to similar precoding vectors, resulting in limited angular resolution.
In the high-overhead regime ($\lambda = 100$), the precoding vectors organize into a smooth, continuous manifold. Neighboring points correspond to gradually varying beamforming strategies, while the dominant AoD changes smoothly along the embedding. Therefore, the precoder representation preserves the dominant-AoD ordering, enabling more adaptive and spatially selective beamforming across propagation conditions. 
These observations reveal that the learned precoder undergoes a transition similar to that observed in the latent feedback space. Under strong compression constraints, the decoder adopts a discrete and highly compressed beamforming strategy that prioritizes robustness and reduced overhead. As the feedback budget increases, the decoder progressively shifts toward a geometry-aware precoding function capable of producing highly adaptive and spatially selective beamforming patterns.
In summary, this analysis demonstrates that the proposed framework does not merely compress CSI information, but instead learns a structured mapping between channel geometry and beamforming behavior, dynamically adapting its spatial resolution according to the available feedback resources.

\begin{figure*}[t]
\centering
\resizebox{\textwidth}{!}{
    \includegraphics[trim={0.2cm 0.0cm 0.2cm 0.0cm},clip,width=\textwidth]{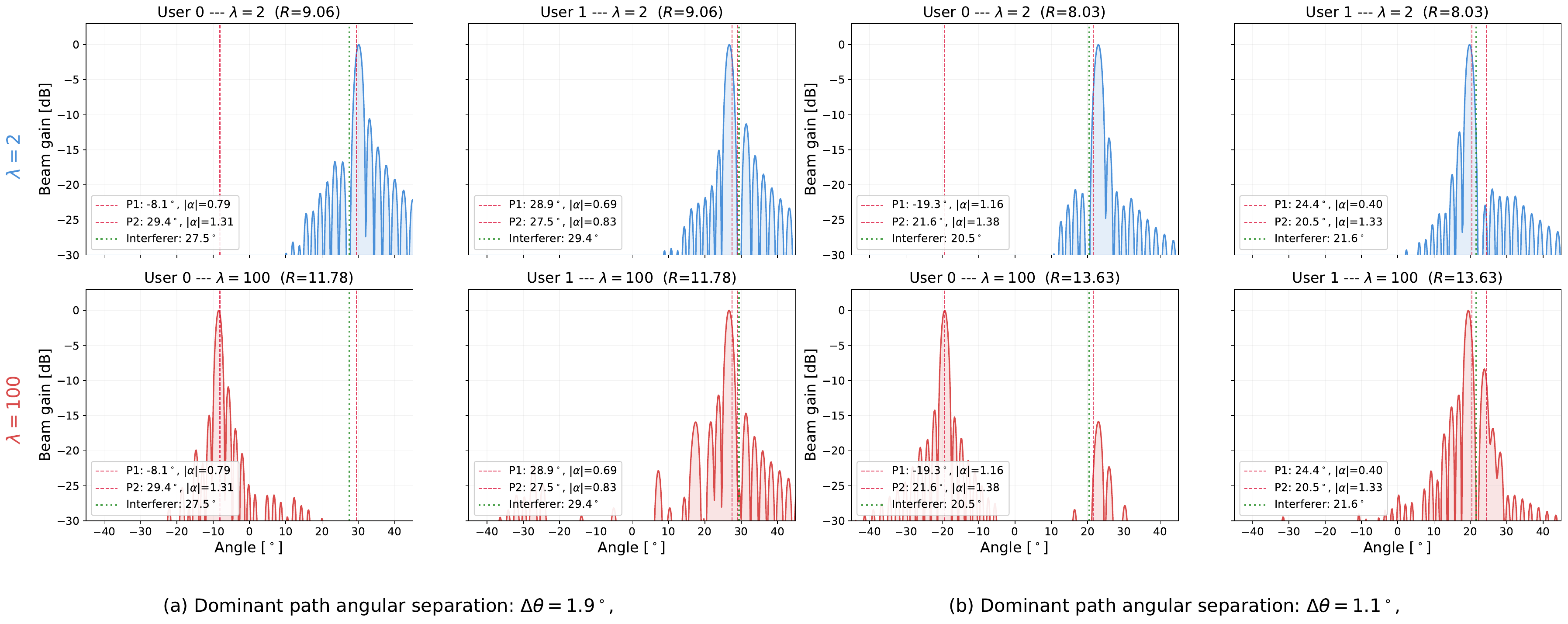}
}
\caption{Beam patterns of the learned precoders for two close-user channel realizations under $\lambda = 2$ (top row) and $\lambda = 100$ (bottom row).
Red dashed lines indicate the user's own multipath AoDs with corresponding path gains $|\alpha|$; green dotted lines indicate the interfering user's
    dominant AoD. P1 and P2 denote path 1 and 2 of the stated user.}
\label{fig:beams}
\end{figure*}

\subsubsection{Beam Pattern Analysis and Interference Management}

To evaluate the impact of the feedback constraint on the learned precoding strategy, we analyze the beam patterns produced by models trained with different values of $\lambda$. 

Fig.~\ref{fig:beams0} presents representative beam patterns  ($\lambda = 2$ on top, $\lambda = 100$ at bottom) for well-separated users ($\Delta\theta \approx 35^\circ$--$42^\circ$) 
selected to have achievable rates close to the average for their respective $\lambda$ values.
Because inter-user interference is naturally low, both models successfully direct energy toward the dominant paths. However, a clear difference emerges in multipath utilization: for $\lambda = 2$, the BS forms a single dominant beam aligned with the strongest path, placing little energy on the weaker path. Conversely, for $\lambda = 100$, the precoder develops dual-beam structures that distribute energy across multiple paths. Thus, in well-separated scenarios, larger feedback primarily enables finer multipath-aware beam shaping rather than strict interference avoidance.

The impact of the feedback constraint becomes pronounced in the closely spaced user scenarios shown in Fig.~\ref{fig:beams}. In Fig.~\ref{fig:beams}(a), the dominant paths of the two users are separated by only $\Delta\theta = 1.9^\circ$. For $\lambda = 2$, the precoder for User~0 concentrates energy around its strongest path ($29.4^\circ$), which nearly overlaps with User 1’s dominant direction ($27.5^\circ$), causing substantial inter-user interference. In contrast, for $\lambda = 100$, the learned precoder redirects energy toward a secondary path at $-8.1^\circ$, avoiding the congested angular region and improving the sum-rate. 
A similar behavior is observed in Fig.~\ref{fig:beams}(b), where the angular separation is further reduced to $\Delta\theta = 1.1^\circ$. 
The low-$\lambda$ model again concentrates energy on the strongest path, whereas the high-$\lambda$ model redirects energy toward the alternative path at $-19.3^\circ$, thereby improving spatial separation and reducing interference.

These observations confirm two distinct precoding behaviors learned by the network:

\begin{itemize}
\item \textbf{Gain-driven strategy ($\lambda$ small):} The precoder prioritizes the strongest channel paths and maximizes beamforming gain, even when this causes inter-user interference.

\item \textbf{Interference-aware strategy ($\lambda$ large):} The precoder exploits additional spatial degrees of freedom and alternative propagation paths to improve user separation, even at the expense of reduced instantaneous channel gain.
\end{itemize}

The transition between these regimes demonstrates that the feedback budget directly influences the learned spatial processing: limited feedback forces the model to maximize raw signal strength, whereas larger budgets enable more sophisticated, end-to-end interference-management.

Our interpretability conclusions are summarized as follows:

{
\renewcommand{\labelitemi}{$\triangleright$}
\begin{itemize}
    \setlength{\itemsep}{0pt}
    \setlength{\parskip}{0pt}
    \setlength{\parsep}{0pt}

    \item  Our interpretations through coloring of t-SNE plots of the precoding vectors indicate that the precoder space transitions from clustered, codebook-like behavior under limited feedback to a smooth, geometry-aware representation at higher feedback budgets.

    \item For well-separated users, by comparing learned precoder beampatterns, it appears that additional feedback enables finer multipath-aware beam shaping and better utilization of secondary paths.

    \item For closely spaced users, the learned strategy shifts from gain-driven beamforming to interference-aware precoding that exploits alternative propagation paths to improve user separation.
\end{itemize}
}

\subsection{Interpretation Under Increasing User Density}
To evaluate the scalability of the learned precoding strategy, we increase the number of users from $K=2$ to $K=4$ while keeping all remaining system parameters unchanged. We then analyze the average received power matrix
$\mathbb{E}[|\mathbf{h}_j^H \mathbf{v}_k|^2]$, where diagonal entries represent desired signal power and off-diagonal entries correspond to inter-user interference.
\begin{figure}[t]
\centering
\resizebox{\columnwidth}{!}{
    \includegraphics[trim={0.2cm 0.0cm 0.2cm 0.0cm},clip,width=\textwidth]{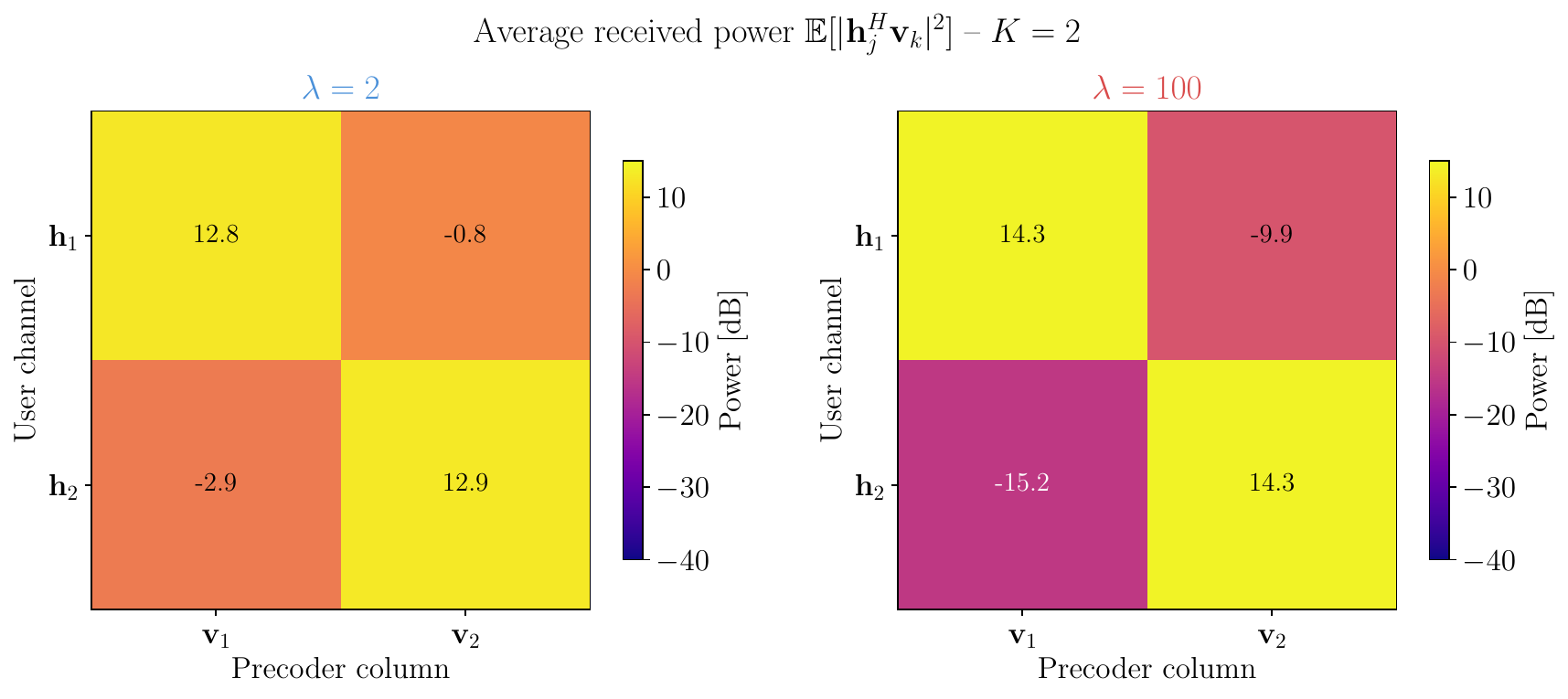}
}
  \caption{Average received power matrix $\mathbb{E}[|\mathbf{h}_j^H \mathbf{v}_k|^2]$ for $K=2$ users.}
\label{fig:interK2}
\end{figure}

\begin{figure}[t]
\centering
\resizebox{\columnwidth}{!}{
    \includegraphics[trim={0.2cm 0.0cm 0.2cm 0.0cm},clip,width=\textwidth]{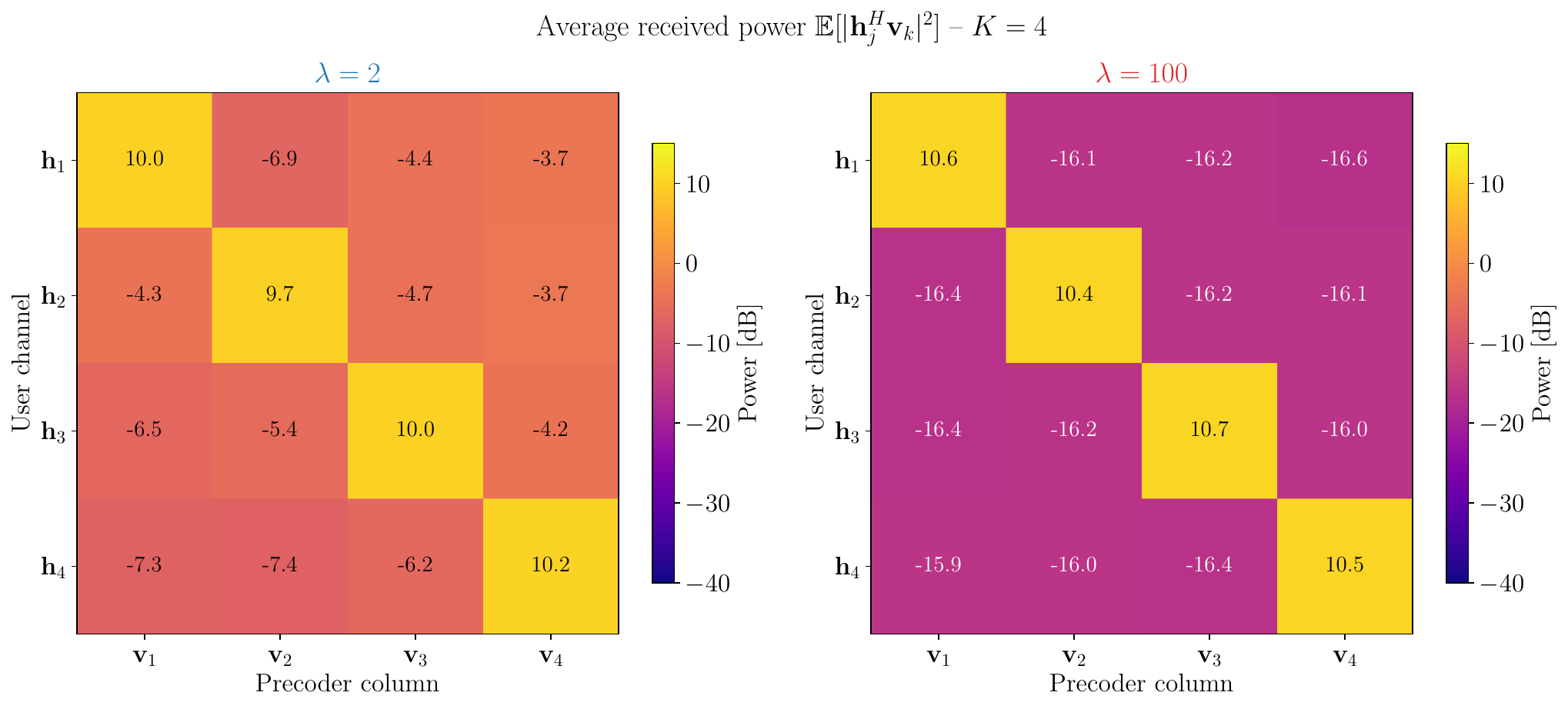}
}
  \caption{Average received power matrix $\mathbb{E}[|\mathbf{h}_j^H \mathbf{v}_k|^2]$ for $K=4$ users.}
\label{fig:interK4}
\end{figure}

Fig.~\ref{fig:interK2} illustrates the learned interference behavior for the two-user case. For $\lambda=2$, the precoder maintains strong desired signal power but exhibits limited interference suppression capability, as reflected by the relatively strong off-diagonal components. In contrast, for $\lambda=100$, the off-diagonal interference terms are significantly reduced while preserving high signal power on the diagonal entries, indicating that the learned precoder adopts a more interference-aware beamforming strategy when larger feedback resources are available.
The same behavior can be observed more clearly in the four-user scenario shown in Fig.~\ref{fig:interK4}. As the number of users increases, interference management becomes substantially more challenging because each beam must simultaneously avoid multiple unintended receivers. Consequently, the low-feedback regime ($\lambda=2$) exhibits noticeable inter-user interference despite maintaining strong received signal power. In contrast, the high-feedback regime ($\lambda=100$) continues to effectively suppress interference across all users. 

Although increasing the number of users naturally reduces the achievable per-user signal power due to spatial resource sharing, the learned precoder maintains a comparable SINR gap of approximately $27$~dB for $\lambda=100$. This demonstrates that the proposed framework scales effectively with the system dimension and that the learned interference suppression mechanism remains robust even as the multi-user precoding task becomes increasingly complex.

Our interpretability conclusions are summarized as follows:
{ \renewcommand{\labelitemi}{$\triangleright$} 

\begin{itemize} 
\setlength{\itemsep}{0pt} 
\setlength{\parskip}{0pt} 
\setlength{\parsep}{0pt} 
\item Interference management becomes more challenging as the number of users increases. 
\item With sufficient feedback, the learned precoder maintains strong interference suppression in both user settings. \end{itemize} }

\section{Conclusion and Future Work}\label{section_6}

This paper presented a comprehensive interpretability analysis of an end-to-end, precoding-oriented CSI feedback framework for FDD massive MIMO systems. While such learning-based designs have recently demonstrated near-optimal downlink performance with significantly reduced feedback overhead, their internal operation has largely remained opaque. To address this limitation, we introduced an XAI–driven methodology to systematically analyze the internal representations and decision mechanisms learned by the joint pilot design, user-side compression, and BS precoding networks.
By combining nonlinear manifold visualization, quantitative density-based clustering, and contingency analysis, we demonstrated that the learned strategy of the end-to-end system is not fixed, but instead emerges as a direct function of the rate–overhead trade-off parameter. In low-overhead regimes, the network resolves the trade-off by developing a structured, semantic compression mechanism that partitions the continuous channel space into a finite set of decision regions. 
This behavior manifests as stable clustering in both the latent feedback representations and the resulting precoding vectors, effectively yielding a soft codebook-like compression strategy, where channels requiring similar transmission decisions are mapped to nearby latent representations and produce similar precoding solutions.
In contrast, as the feedback constraint is relaxed, this discrete structure dissolves, and the system transitions to a continuous, reconstruction-oriented operating regime, where the learned representations are no longer clusterable and closely resemble classical channel estimation behavior.
Importantly, our analysis confirms that the observed structure is not an inherent property of the channel statistics, but a learned response to the imposed feedback constraint. The BS implicitly adopts a hybrid decision strategy: deterministic mappings for unambiguous channel states under tight feedback budgets, and context-aware mappings for more complex interference scenarios. These findings provide concrete evidence that end-to-end learning frameworks internalize meaningful, interpretable strategies to balance compression efficiency and precoding performance.


Beyond interpretation, this work opens several promising research directions into the nature of end-to-end learned communications. A critical open question concerns the exact physical interpretation of the continuous latent representations learned in high-rate regimes. Given that the learned pilots tend to converge to an orthogonal basis (a classical optimality condition for channel estimation) future work could investigate whether the latent space prioritizes faithful channel waveform reconstruction or effectively decouples explicit physical parameters such as path gains and angles. Furthermore, the analytical approximation of the learned mapping from latent vectors to precoding decisions remains a highly promising avenue. Specifically, research should explore whether the network implicitly approximates an adaptive Regularized ZF (RZF) strategy, $V = H^H(HH^H + \alpha I)^{-1}$, by dynamically tuning the regularization parameter $\alpha$ to shift between gain-driven, i.e., MRT, and interference minimization, i.e., ZF, regimes based on the detected spatial conditions. 

Additionally, moving from implicit discovery to explicit control offers significant potential; introducing targeted regularization could force the continuous latent space to collapse into an exact, controllable number of precoding prototypes. Finally, evaluating how these learned codebooks and continuous manifolds adapt to different propagation physics, such as non-sparse, rich-scattering Rayleigh environments, would be critical to assessing the universal applicability of these architectures.

\bibliographystyle{ieeetr}
\bibliography{ref}
\end{document}